\documentstyle{article}
\input epsf
\begin{document}
\rightline{EFI-96-19}
\rightline{hep-th/9605223}
\vskip 1cm
\centerline{\LARGE \bf The target space geometry of}
\centerline{\LARGE \bf $N=(2,1)$ string theory}
\vskip 1cm

\renewcommand{\thefootnote}{\fnsymbol{footnote}}
\centerline{{\bf \large Albion Lawrence\footnote{
Supported in part by Dept. of Energy grant
DEFG02-90ER-40560}\footnote{Address after Sept. 1, 1996: 
Lyman Laboratory
of Physics, Harvard University, Cambridge, MA, 02138}}}
\vskip .5cm
\centerline{\it Enrico Fermi Inst. and Dept. of Physics}
\centerline{\it University of Chicago,
5640 S. Ellis Ave., Chicago, IL 60637 USA}
\centerline{\it albion@yukawa.uchicago.edu}
\setcounter{footnote}{0}
\renewcommand{\thefootnote}{\arabic{footnote}}

\begin{abstract}

We describe the ${\cal{O}}({\alpha'}^0)$ constraints on the 
target space geometry of the $N=(2,1)$ heterotic superstring
due to the left-moving $N=1$ supersymmetry and 
$U(1)$ currents.  In the fermionic description of the internal sector
supersymmetry is realized quantum mechanically, so that both
tree-level and one-loop effects contribute to the order
${\cal{O}}({\alpha'}^0)$ constraints.  
We also discuss the physical interpretation of the resulting target
space geometry in terms of configurations of a
$2+2$-dimensional object propagating in a $10+2$-dimensional
spacetime with a null isometry, which has recently been suggested
as a unified description of string and M theory.

\end{abstract}
%

\def\figloc#1#2{\bigskip\vbox{{\epsfxsize=4.5in
        \nopagebreak[3]
    \centerline{\epsfbox{diag#1.ps}}
        \nopagebreak[3]
    \centerline{Figure #1}
        \nopagebreak[3]
   {\raggedright\it \vbox{  #2 }}}}
    \bigskip
    }

\def\pref#1{(\ref{#1})}

\def\ie{{\it i.e.}}
\def\eg{{\it e.g.}}
\def\cf{{\it c.f.}}
\def\etal{{\it et.al.}}
\def\etc{{\it etc.}}

\def\adv{{\it Adv. Phys.}}
\def\ap{{\it Ann. Phys, NY}}
\def\cqg{{\it Class. Quant. Grav.}}
\def\cmp{{\it Commun. Math. Phys.}}
\def\jetp{{\it Sov. Phys. JETP}}
\def\jetpl{{\it JETP Lett.}}
\def\jp{{\it J. Phys.}}
\def\ijmp{{\it Int. J. Mod. Phys. }}
\def\nc{{\it Nuovo Cimento}}
\def\np{{\it Nucl. Phys.}}
\def\mpl{{\it Mod. Phys. Lett.}}
\def\pl{{\it Phys. Lett.}}
\def\pr{{\it Phys. Rev.}}
\def\prl{{\it Phys. Rev. Lett.}}
\def\prcl{{\it Proc. Roy. Soc.} (London)}
\def\rmp{{\it Rev. Mod. Phys.}}
\def\dash{----------    }

\def\p{\partial}

\def\apr{\alpha'}
\def\str{{str}}
\def\lstr{\ell_\str}
\def\gstr{g_\str}
\def\Mstr{M_\str}
\def\lpl{\ell_{pl}}
\def\Mpl{M_{pl}}
\def\varep{\varepsilon}
\def\del{\nabla}
\def\tr{\hbox{tr}}
\def\perp{\bot}

\def\AA{{\cal A}}
\def\BB{{\cal B}}
\def\CC{{\cal C}}
\def\DD{{\cal D}}
\def\EE{{\cal E}}
\def\FF{{\cal F}}
\def\GG{{\cal G}}
\def\HH{{\cal H}}
\def\II{{\cal I}}
\def\JJ{{\cal J}}
\def\KK{{\cal K}}
\def\LL{{\cal L}}
\def\MM{{\cal M}}
\def\NN{{\cal N}}
\def\OO{{\cal O}}
\def\PP{{\cal P}}
\def\QQ{{\cal Q}}
\def\RR{{\cal R}}
\def\SS{{\cal S}}
\def\TT{{\cal T}}
\def\UU{{\cal U}}
\def\VV{{\cal V}}
\def\WW{{\cal W}}
\def\XX{{\cal X}}
\def\YY{{\cal Y}}
\def\ZZ{{\cal Z}}

\section{Introduction}

The suspicion that the fundamental, short distance
description of string theory may
not involve strings has existed for a number of years, 
based on the theory's 
behavior at high energies and at large orders in perturbation
theory (for a
review of these arguments with references, see 
the final two chapters of Polchinski (1994)).
In addition, developments in string duality (For a recent review
and references see Hull (1995)) have provided a picture of the vacuum
structure of string theory which casts some doubts on its
status as a fundamental theory.  At various extremes of the moduli
space of vacua a given string theory may have a weakly coupled 
description as a different string theory, as 11-dimensional supergravity,
as ``M-theory'' (Schwarz 1995b,c), or as some other p-brane theory
(note however that Hull (1995) has suggested that the perturbative
states of the supersymmetric effective field theories are always strings
or $0$-branes).  Furthermore, p-branes wrapped around
homology cycles of the compactification manifold sometimes 
turn into string states when one condenses them, as happens in
conifold transitions 
(Greene, Morrison and
Strominger 1995).  Wrapped $p$-brane states may
also be related by duality to
string states (for a recent review and discussion see Townsend 1995).
This has led some to suggest there may be a more ``democratic'' theory in
which strings are treated on the same footing as other $p$-brane
excitations (Becker, Becker and Strominger 1995; Townsend 1995).

Still, there is little indication as to what the underlying theory might be.
M-theory has no definition
beyond its being an 11-dimensional theory with supersymmetric 2-branes;
the quantum theory of 11-dimensional supergravity is unknown; 
and the analog of the Polyakov action for p-branes
with $p>1$ is nonrenormalizable, partly due to the coupling to
worldvolume gravity (for a discussion of this point see 
van Nieuwenhuizen 1987).  Furthermore, one could argue that these
dualities are merely features of the low energy theory,
and that one particular object
defines the short distance theory.
As an analogy, one may write a theory of strongly coupled
electrons in a one-dimensional metal, described by the massive Thirring model.
There will be a weakly coupled description in terms of the
Sine-Gordon model (Coleman 1975), but the theory defined
at the lattice scale really is a theory of electrons.
So in the case of ``string'' theory there seems to be no reason to
privilege any one $p$-brane state as fundamental,
but there is no obvious reason not to,
beyond the high energy and large-genus problems alluded to above.

One possibility for a more fundamental description of
string theory is suggested by some
recurring hints that the worldsheet might be
secretly
four dimensional, propagating in a spacetime with more
than one timelike signature.  Atick and Witten (1988) suggested
that the time direction might be complexified, based on an
attempt to interpret the high-temperature behavior of string theory.
Witten (1988) suggested that both the worldsheet and spacetime
might be secretly complexified.  This was based on a comparison 
of instantons in an orbifolded 
topological $\sigma$-model with
classical configurations dominating high energy string scattering processes
(Gross and Mende 1987,1988;Polchinski 1988).
Blencowe and Duff (1988)
conjectured that the maximum possible spacetime dimensions
for a $p$-brane theory
were 12; this arose from demanding
equal Bose and Fermi degrees of freedom
on the worldvolume, while allowing for a spacetime supersymmetry group 
other than the super-Poincar\'{e} group.  As an example they mentioned
a $2+2$ dimensional object in a $10+2$ dimensional spacetime
that 
might be related to the Type $IIB$ string
by double dimensional reduction.
More recently, Hull (1995) has suggested an
$11+1$-dimensional ``Y-theory'' in order to explain certain
6-dimensional supergravity theories with exotic
soliton spectra.

Still more recently, Kutasov and Martinec (1996) have found that the
bosonic string in 26 dimensions; the heterotic and type IIB strings
in 10 dimensions; a bosonic
27 dimensions; and a supermembrane in 11 dimensions
can all be found as
different vacua of the $N=(2,1)$ supersymmetric heterotic
string constructed by Ooguri and Vafa (1991b).
The open string and  
examples involving orbifold
and orientifold compactifications of string- and M-theory 
have been found found by Kutasov,
Martinec and O'Loughlin (1996).  In these constructions the target
space is in fact a $2+1$-dimensional object, with a $2+2$-dimensional
worldvolume, living in a flat
$10+2$-dimensional (or $26+2$-dimensional if we give up
target space supersymmetry);
one or two of these dimensions are gauged
away by a left-moving timelike or null $U(1)$ current on the 
worldsheet of the $N=(2,1)$ string.  Which string or membrane
theory one gets depends on how one realizes
the left-moving $U(1)$ and how one performs the $GSO$ projection.
The target space
theory has a lot of structure: it seems to be
self-dual gravity coupled to self-dual Yang-Mills fields, in the presence
of a covariantly conserved timelike or null Killing vector
(Ooguri and Vafa 1991b; see also Pierce 1996).  This might be enough structure
to allow the $2+1$-dimensional object 
to be quantized directly.

Simultaneously, Vafa (1996) has suggested
a $10+2$-dimensional ``F-theory'' in order to
explain the 
$SL(2,\ZZ)$ duality of type IIB string theory (Schwarz 1995a)
directly in 10 dimensions; he also suggested that this was required
by
gauge fields living on the worldsheets of the D-strings of IIB 
string theory.  The $SL(2,\ZZ)$ duality suggests a hidden torus whose
modular group is precisely this duality group, and the gauge
field on the D-string
requires an increase of 2 in the critical dimension due to the
$U(1)$ worldsheet ghosts.  Vafa further suggested
that a geometrization of the $U(1)$ algebra would add an additional
two dimensions to the worldsheet, though how this comes about and how
these extra two dimensions are mapped into spacetime is unknown.
Tseytlin (1996) also simultaneously
proposed a $11+1$ dimensional theory
of Dirichlet $3$-branes (with a worldvolume signature of $(3,1)$), 
based on an interpretation of the gauge fields living on
the self-dual $3$-brane of type IIB string theory.

Given this small explosion of suggestions of a theory of 
4-di\-men\-sion\-al
worldvolumes in 12 dimensions, 
we would like to begin to understand what precisely
this theory might be.  The $N=(2,1)$ string gives us a solid starting
point; it has all known string theories and
$M$-theory as different vacua, and it should have a definite target
space theory which would describe the dynamics (whatever
this means in the presence of 2 timelike directions)
of these 3-dimensional objects.
However, computations from the 
point of view of the $(2,1)$ string worldsheet seem rather unwieldy.
The vertex operators of the underlying string theory describe at best
linearized fluctuations of the target worldvolume; and describing the
quantum mechanics of the worldvolume requires the string field theory of
the underlying $(2,1)$ string.

In this paper we will start to examine the target space theory
by understanding the constraints on the $\sigma$-model geometry of
the $N=(2,1)$ string due to the worldsheet supersymmetry and left-moving
$U(1)$ current algebra. Our attitude is that while the $(2,1)$ string
may be unwieldy for describing the quantum mechanics of the theory,
one should use whatever structure is available to gain
insight; in particular, an understanding of which fields the
$(2,1)$ string can couple to 
should lead to some
understanding of the configuration space of the $3$-brane theory.
We will take a very small step in this direction, by
deriving to $\OO(\apr^{0})$ the constraints on the target space
fields due to the $(2,1)$ worldsheet supersymmetry and the
additional left-moving $U(1)$ symmetry.
Such constraints have been discussed previously (Hull 1986b, Dine
and Seiberg 1986, Braden 1987);
in all of these discussions, the internal
sector of the heterotic string was described by left-moving fermions 
(as these have the most natural coupling to the gauge fields) and
the left-moving supersymmetry was imposed at tree level.  These authors
thus found that
was that the internal fermions did not propagate and the gauge field
was forced to be trivial.  However, it is known that supersymmetry
may be nontrivially realized on free fermions;
in this case the algebra closes only at one loop (Goddard, Nahm and Olive
1985; Goddard, Kent and Olive 1986; Windey 1986;
Antoniadis \etal\ 1986; Gates, Howe and Hull 1989).  In order to
get the correct coefficients of the current commutators, the internal
supersymmetry currents must scale as $1/\sqrt{\apr}$; therefore
one-loop terms
will appear at order $\OO(\apr^{0})$ in the commutator
of the left-moving supersymmetry current $G_{-}$ with itself.
We will also find that $\OO(\sqrt{\apr})$ 
corrections to the internal gauge fields (Equation \pref{conncorr})
and to the supersymmetry current (Equation \pref{fcorr})
mix with the internal supersymmetry currents to produce
terms in $\{G_{-},G_{-}\}$ at $\OO(\apr^{0})$. 
The result is that the internal theory can in fact be nontrivial
in the presence of a left-moving superconformal symmetry.

The left-moving fermionic sector of the theory gives
a nice geometrical structure (see also Braden (1987)).
One may describe the left-moving fields of the $N=(2,1)$ string with 
4 bosonic worldvolume coordinates and 28 fermions living in 
some vector bundle $\VV$ fibered over the worldvolume.  It is only
after imposing supersymmetry that 4 of
these fermions are associated with the left-moving excitations of the
worldvolume coordinates and the gauge curvature in
these directions is set equal to the worldvolume curvature.  
The identification of the tangent
plane of the worldvolume can and will
vary fiber by fiber in $\VV$, as specified by the
supersymmetry current; this identification will split $\VV$ into
``tangent'' and ``normal'' bundles.  Furthermore, the
fibers of the ``normal'' bundle must live in the adjoint
representation of some Lie group (which may be a product group);
this group structure is encoded by the left-moving supersymmetry current
as well.  Thus the supersymmetry current adds some additional
structure.  Constraints on this structure
and on the $\sigma$-model geometry are found in Equations \pref{esquared},
\pref{feqh},\pref{jacobi},\pref{cartanmetric}, \pref{cove}, and
\pref{reqf2}.  If we interpret the target space theory as
describing the imbedding of a $2+2$-dimensional object in some
spacetime, $\VV$ should be related (nontrivially, as we
will argue) to the spacetime.

More structure will arise as
the $N=(2,1)$ string has a gauged, left-moving $U(1)$ 
supercurrent.  We will find that the piece
of the $U(1)$ current acting on the internal fermions must
scale as $1/\sqrt{\apr}$, and so as with the supersymmetry
algebra we will find that one-loop terms arise at
order $\OO(\apr^{0})$, and that we will
have to add a
piece scaling as $\sqrt{\apr}$ to the supercurrent
(Equation \pref{sigmaimprove}) in order to maintain consistency.
The constraints on the geometry and the supercurrent
to $\OO(\apr^{0})$ are shown in Equations \pref{lieg},\pref{lieb},
\pref{delminusv}, and \pref{schwingvanish}.  The upshot is that
the target space possesses a null or timelike covariantly
conserved Killing vector; if it is timelike, there must
be an $\OO(1/\sqrt{\apr})$ piece of the supercurrent acting on the
internal fermions to cancel the anomaly of the target space isometry.

This thesis is organized as follows.
Section 2 is a review of known pertinent results.  
Section 2.1 reviews the
properties of $N=(2,1)$ strings with a flat target space; it
also reviews the realization of supersymmetry on free fermions.  Section
2.2 reviews the general $\sigma$-model action for the heterotic
string and the constraints arising from $(2,0)$ supersymmetry as derived
by Hull and Witten (1985).  Section 3 describes the 
left-moving supersymmetry and $U(1)$ currents; it also
lists the transformations
of the fundamental $\sigma$-model fields under Dirac brackets
with these currents.  Section 4 contains the derivation of
the constraints on the $\sigma$-model geometry arising from
the left-moving supersymmetry.
Section 5 contains the constraints due to the left-moving $U(1)$
supercurrent.
Section 6 discusses of the physical and geometric
interpretation of our results in terms of a 4-dimensional surface
living in some spacetime.  Section 7 contains conclusions and 
a discussion of directions
for further work.  Appendix A contains conventions for worldsheet and
target space geometry. Appendix B describes the Dirac
brackets of the $N=(2,1)$ $\sigma$-model.  
Finally, Section C contains conventions
for the worldsheet Green's functions.  

A note about nomenclature: in this paper the {\it right}-moving sector of
the $N=(2,1)$ string will be the $N=2$ sector and the {\em left}-moving
sector will be the $N=1$ sector.  This is the
convention used by Kutasov and Martinec (1996) and 
Kutasov, Martinec and O'Loughlin (1996).

\section{Background}

\subsection{The $N=(2,1)$ string in flat spacetime}

In order to get a feeling for the $N=(2,1)$ string we will first review
the flat space theory.  Many features of the $\sigma$-model in
arbitrary backgrounds will be generalizations of this simple case;
also, the physical massless states 
of the theory should tell us which backgrounds to turn on.

Strings with gauged $N=(2,2)$ worldsheet supersymmetry have 
been studied by Ademollo \etal\ (1976a,b);
Fradkin and Tseytlin (1981, 1985a); D'Adda and Lizzi (1987); Green (1987);
and Mathur and Mukhi (1987, 1988).  Ooguri and Vafa (1990, 1991a)
explored the target space geometry more systematically by exploring
the scattering amplitudes for physical states and the target space effective
action one could deduce from these amplitudes.  They also constructed
heterotic string theories 
with gauged $(2,0)$ and $(2,1)$ supersymmetry (1990,1991b);\footnote{Fradkin 
and Tseytlin (1985a) and Green (1987) 
also described the $(2,0)$ and $(2,1)$
heterotic string.  In both papers the necessity of the
left-moving $U(1)$ current was missed, although in the latter a truncation
to a 2-dimensional target space was assumed.}
we will review these constructions here.

The right-moving sector of these theories includes in addition 
to the reparameterization
and supersymmetry ghosts a pair of ghosts corresponding to the
$U(1)$ piece of the $N=2$ algebra (the complex structure).  The critical
dimension of the right-movers is 4, and $N=2$ SUSY requires
a complex structure (Alvarez-Gaum\'{e} and Freedman 1981;
Hull and Witten 1985),
so we must have either $(4+0)$ or $(2+2)$ signature.
We will choose the latter.  The left-moving bosons
must then also have two timelike directions (in particular, if
the target space is noncompact); however, the $N=1$ superconformal
symmetry will only remove negative-norm states
arising from one timelike direction.  Unitarity
requires an additional left-moving $U(1)$ gauge symmetry in order to remove
the remaining negative-norm states.
This gauging increases the critical dimension of the
left-moving sector by 2, as there will be Faddev-Popov ghosts
associated with the $U(1)$.
Thus the matter sector of the left-movers of the 
$(2,0)$ string has $c=28$; the matter sector of the left-movers of the
$N=(2,1)$ string has $\hat{c}=12$.\footnote{$\hat{c}=1$ for
a free chiral superfield, or equivalently for three free Majorana-
Weyl fermions.}

Since the left-moving $U(1)$ current is gauged, nilpotency of the
BRST charge requires
that this current must have a vanishing operator product
with itself (Ooguri and Vafa 1991b).  
In the case of the $N=(2,0)$ string
in flat space, the $U(1)$ current will have the form
\begin{equation}
J_{-} = v_{\mu}\p_{-}\phi_{L}^{\mu}\ ,
\end{equation}
where $\phi_{L}$ denote both the left-moving target worldvolume coordinates
and the chiral scalars of the internal sector.  It is easy to
see that the condition for the vanishing OPE of this
current with itself is
$v_{\mu}v^{\mu} = 0$.

The internal sector of the
$N=(2,0)$ string can be represented by 24 chiral scalars which live
in the maximal torus of a rank 24 group.
The physical states of the
theory depend on the direction of the left-moving $U(1)$, i.e. on
whether the vector has any components in the internal direction.
With no components of the $U(1)$ in the internal direction the null
isometry in $2+2$ dimensions kills off the gravitational
dynamics of the target worldvolume
gravitational dynamics, so that the only excitations 
come from the internal sector
and correspond to the gauge bosons of the internal gauge group.
Scattering amplitudes indicate that the target spacetime theory is
the self-dual Yang-Mills theory in $2+2$ dimensions, reduced to
a $1+1$-dimensional theory by the isometry.  
If the $U(1)$ has components in the internal 
direction, then the spacetime isometry is timelike and so the target
spacetime is effectively $2+1$-dimensional. In this case there is some
spacetime boson which is a remnant of the boson describing
self-dual gravity in the $(2,2)$ string (Ooguri and Vafa 1990, 1991a).
There are also the internal gauge fields as before;
the gauge symmetry is partly broken by the internal part of
the $U(1)$ (Ooguri and Vafa 1991b).  In this case the theory
describes some sort of coupling of self-dual Yang-Mills and self-dual
gravity.

For the $N=(2,1)$
string Ooguri and Vafa represented the internal sector by
eight chiral scalars and
their fermionic superpartners.  They found that the only massless 
states correspond to the 8-dimensional ground state of the Ramond 
sector of the internal fermions and to the 8 
Neveu-Schwarz states created by single fermion oscillators acting on the
vacuum.
Ooguri and Vafa therefore claimed that the internal sector
has no group structure.  However, one can find other realizations
of the internal sector which do contain gauge symmetry.
For example, Pierce (1996) used a free fermion
construction of the internal SCFT, and found that the massless states
in the adjoint of a 24-dimensional Lie algebra
with the fermions living in the adjoint of this group.  As we will
use the fermionic representation of the $(2,1)$ $\sigma$-model below,
let us review supersymmetric theories of free Majorana-Weyl fermions.

Given a theory with 
free left-moving Majorana-Weyl fermions $\lambda^{a}$, 
one may realize an affine Lie algebra with fermion bilinears
(for a review and references see Goddard and Olive 1986).
If we let the fermions live in the adjoint representation of a Lie group $H$,
one may realize a worldsheet current algebra corresponding to this group:
\begin{equation}
J^{a} = f^{abc}\lambda^{b}\lambda^{c}\ ,
\end{equation}
where the coefficients $f^{abc}$ are the structure constants of the $H$.
There is a well-known $N=1$ supersymmetry in such theories
(Goddard, Nahm and Olive 1985; Goddard, Kent and Olive 1986; 
Windey 1986; Antoniadis \etal\ 1986), where
the supersymmetry charge is
\begin{equation}
G = \frac{1}{3 \sqrt{C_{A}}} f^{abc}\lambda^{a}\lambda^{b}\lambda^{c}\ .
\label{fermsusy}
\end{equation}
This realization with fermions living in the adjoint of a group
is general.
Following Windey (1986) and Antoniadis \etal\ (1986)
we can realize $N=1$ supersymmetry on a set of free Majorana-Weyl
fermions by splitting them up into fermions living in the
adjoint of some group $H$ and fermions living in some representation of
that group.  Requiring that the superconformal algebra
closes appropriately means that sum of these spaces
is in fact the adjoint representation of some group
$G\supset H$ and that $G/H$ is a homogeneous space.  The level of the current
algebra is the dual Coxeter number of this group.  Ooguri and
Vafa (1991b) used the vertex operator construction of the
current algebra (Halpern 1975; Frenkel and Kac 1980; Segal 1981; 
Gross \etal\ 1985, 1986) which is a level 1 construction.
If we fermionized this theory we would find that the only
current algebra compatible with left-moving supersymmetry at this level
is $U(1)^{8}$.

If we wish to find which states correspond to
gauge bosons and what the gauge structure is, we need to choose an
appropriate GSO projection compatible with modular invariance.
The spectrum of massless states and its group structure depends on the
choice of SUSY current and on
the GSO projection (Pierce 1996), just as in the free fermion models
of type II string
theories with gauge symmetry (Kawai, Llewellyn and Tye 1986a,b and 1987a,b; 
Bluhm, Dolan and Goddard 1987).  In particular the construction of
different string and membrane theories as target spaces of the $(2,1)$ string
relies on the spectra one gets from different GSO projections
(Kutasov and Martinec 1996).

It is also worth recalling the conditions for target space supersymmetry.
Since the right-moving part of the theory has
gauged $N=2$ supersymmetry, the spectral flow is gauged; thus
the Neveu-Schwarz and Ramond sectors of the right-movers
are equivalent (Schwimmer and Seiberg 1987; Ooguri and Vafa 1991b).
Spacetime supersymmetry must arise entirely from the
left-movers; however,
a well-known theorem of Friedan and Shenker (unpublished; for
a published description and discussion see Section 2.4 of 
Dixon, Kaplunovsky and Vafa (1987)) states that target space
supersymmetry and non-Abelian gauge symmetry cannot arise
from the same sector.  
Equivalently, target space supersymmetry
exists if and only if the left-moving sector possesses global
$N=2$ supersymmetry (Banks \etal\ 1988; Banks and Dixon 1988).
Thus, such a global supersymmetry is also incompatible with
non-Abelian gauge symmetry. Indeed, the constructions of
Kutasov and Martinec (1996) and
Kutasov, Martinec and O'Loughlin (1996) which possess spacetime supersymmetry
have no non-Abelian structure.

\subsection{The $\sigma$-model action, 
and constraints from $(2,0)$ supersymmetry}

We are interested in the more general $\sigma$-model which one would find
upon condensing vertex operators corresponding to the 
massless bosonic physical states
of the $(2,1)$ string.  We know that in general the bosonic sector contains
target space gravity and gauge fields.  We will also include an antisymmetric
tensor field, as torsion arises naturally in
heterotic $\sigma$-models: since no antisymmetric tensor field appears in
the physical state spectrum this field will merely
dress the gauge and gravitational excitations in solutions to
the $\beta$-function equations.

We can break up the fields of the $N=(2,1)$ sigma model into several
pieces: $d$ target worldvolume coordinates which we can
write as $N=(2,0)$ superfields, making the right-moving supersymmetry
manifest (Dine and Seiberg, 1986);  
$d$ left-moving Majorana-Weyl fermions
to pair with the left-moving target space bosons; and a piece
describing the
internal left-moving conformal field theory.
The internal theory
can be written as a set of $3n$ real, left-moving fermions with indices
in some $3n$-dimensional tangent space, coupled to
a background vector potential, as a set of $n$ left-moving chiral scalars
paired to $n$ left-moving real fermions by the $N=(0,1)$ SUSY, 
or as any other left-moving $\hat{c}=8$ superconformal field theory;
we will work with the fermionic representation, since
the coupling to target space gauge fields is straightforward.

The action for the target space bosons and their
right-moving fermionic superpartners may be written in
$(2,0)$ superspace following Dine and Seiberg (1986):
\begin{equation}
S_{st} = -\frac{i}{2}\int d^{2}\sigma d\theta d\theta^{\ast} 
	\left[ K_{i}(\Phi^{j}, \Phi^{\bar{k}})
	\p_{-}\Phi^{i} - K_{\bar{i}}(\Phi^{j},\Phi^{\bar{k}})
	\p_{-}\Phi^{\bar{i}} \right] \ .
\label{n2ract}
\end{equation}
where we use the notation given in Appendix A.  
The action in component form is
(Hull and Witten 1985):
\begin{eqnarray}
\lefteqn{S_{st} = \frac{1}{2}\int d^2 \sigma \left[
	\left(g_{\mu\nu}(\phi) 
		+ b_{\mu\nu}(\phi)\right)
	\p_{+}\phi^{\mu}\p_{-}\phi^{\nu}\right. }  \nonumber \\
	&&\left. + i g_{\mu\nu} \psi^{\mu} \left(
	\p_{-} \psi^{\nu} + {\Gamma_{(+)}}^{\nu}_{\lambda\rho}(\phi)
	\p_{-}\phi^{\lambda}\psi^{\rho}\right)\right] \ .
\label{ract}
\end{eqnarray}
We will work with this form, dropping the manifestly complex
parameterization; there will be a complex structure tensor
$J^{\mu}_{\nu}$ with the
properties required for $(2,0)$ supersymmetry, which we will discuss below.

We could write
the left-moving fermions in $(2,0)$ 
superspace as in Dine and Seiberg.  This
would force a Hermitean structure on the vector bundle which the
internal fermions live in, as required for off-shell
closure of the $N=2$ algebra (Howe and Papadopolous 1988).
We will worry only about on-shell closure in this paper.
In $(1,0)$ superspace the action for left-moving
Majorana fermions coupled to a background gauge field is
(Hull and Witten 1985):
\begin{equation}
S_{{\rm Rf}} = - \int d^{2}\sigma d\theta_{+}
	G_{ab}(\Phi)\Lambda^{a}\left(
	D_{+}\Lambda^{b} + A_{\mu}{}^{b}{}_{c}D_{+}\Phi^{\mu}
	\Lambda^{c}\right)\ .
\end{equation}
In this equation $\Phi$ is the $(1,0)$ bosonic superfield as written
in Appendix A.
Integrating over the Grassman coordinate $\theta_{+}$ and
eliminating the auxiliary fields, we find the component form
of the action (Hull and Witten 1985):
\begin{equation}
S_{Lf} = \frac{1}{2}\int d^{2}\sigma \left[
	i G_{ab}(\phi) \lambda^{a} \left(\p_{+} \lambda^{b}
	+ {A_{\mu}}^{b}{}_{c} \p_{+}\phi^{\mu}\lambda^{c}\right)
	+ \frac{1}{2} F_{\mu\nu ab}\psi^{\mu}\psi^{\nu}\lambda^{a}
	\lambda^{b}\right]\ .
\label{cuferm}
\end{equation}
One may rotate away the metric $G$ with
the vielbein fields $\rho^{a}{}_{A}$ and obtain the action
\begin{eqnarray}
\lefteqn{S_{Lf} = \frac{1}{2}\int d^{2}\sigma \left[
	i \eta_{AB}(\phi) \lambda^{A} \left(\p_{+} \lambda^{B}
	+ {\omega_{\mu}}^{B}{}_{C} \p_{+}\phi^{\mu}\lambda^{C}\right)\right.}
		\nonumber \\
	&&\left.
	+ \frac{1}{2} F_{\mu\nu AB}\psi^{\mu}\psi^{\nu}\lambda^{A}
	\lambda^{B}\right]\ .
\label{flferm}
\end{eqnarray}
There will also be a dilaton term appearing appearing in the action at 
$\OO(\apr)$; in conformal gauge this should couple to the
ghosts as described by Banks, Nemeschansky and Sen (1986).  One
may redefine the ghost fields in such a way that the dilaton
does not appear in the action; this transformation
is anomalous, so the BRST current picks up a term proportional
to the dilaton at order $\apr$ (for the superstring version of this
in superspace, see Aldazabal, Hussain and Zhang (1987)).  This
is of too high an order to appear in our $\OO(\apr^{0})$ calculation.

The fermions live in a vector bundle $\VV$
with connection $\hat{A}_{\mu}{}^{a}{}_{b}$
fibered over the target space (Hull and Witten 1985).
We expect the left-moving supersymmetry to relate
some 4-plane in each fiber to the tangent space of the worldvolume;
it should also somehow relate
the ``tangent'' gauge connection to the spin connection
(indeed, we will find that the curvatures agree).
We will not make this identification at the level of the
action; rather, all of the information encoding
this identification will lie in the supersymmetry current.
The advantages of this presentation will become clear when we
discuss the interpretation of target space theory as a 4-dimensional
worldvolume immersed in some spacetime.

The combination of actions \pref{ract} and \pref{cuferm}
is invariant under the supersymmetry transformation
\footnote{In the paper of Hull and Witten (1985)
the sign in the variation of 
$\lambda$ in Equation \pref{n1rlam} is incorrect.  Braden (1987) also uses the
sign opposite that of Equation \pref{n1rlam}, but
in his paper the sign is correct as the action for $\psi$ is defined with
a total sign opposite that of the action \pref{cuferm}.}
\begin{eqnarray}
	\delta \phi^{\mu} &=& \epsilon \psi^{\mu} \label{n1rphi} \\
	\delta \psi^{\mu} &=& i \epsilon \p_{+} \phi^{\mu} \label{n1rpsi} \\
	\delta \lambda^{a} &=& - \epsilon {\hat{A}_{\mu}}{}^{a}{}_{c}
	\psi^{\mu} \lambda^{c} \label{n1rlam}
\end{eqnarray}
and the complex structure transformation
\begin{equation}
	\delta \psi^{\mu} = J^{\mu}{}_{\nu} \psi^{\nu} \ , \ 
	\delta \phi = \delta \lambda = 0\ . \label{rcstruc}
\end{equation}
The second supersymmetry comes by commuting the first supersymmetry with the
complex structure (Hull and Witten 1985).
Invariance of equations \pref{ract} and \pref{cuferm}
under the complex structure rotation \pref{rcstruc} combined
with closure of the $N=2$ algebra place requirements on 
the complex structure (Hull and Witten 1985):
\begin{eqnarray}
&&J^{\mu}_{\beta} J^{\beta}_{\nu} = - \delta^{\mu}_{\nu} \label{jeqi}\\
&& N^{\mu}_{\nu\rho} = J^{\beta}_{\nu} \p_{\left[\beta\right.} 
		J^{\mu}_{\left.\rho\right]}
	- J^{\beta}_{\rho} \p_{\left[\beta\right. } 
		J^{\mu}_{\left. \nu\right]} = 0 \label{nienhuis}\\
&& g_{\alpha\beta}J^{\alpha}_{\mu}J^{\beta}_{\nu} = g_{\mu\nu} \\
&& \del_{(+)\lambda} J^{\mu}_{\nu} = 0 \label{cscons} \label{delcom}\\
&& J^{\mu}{}_{\left[\nu\right.} F_{\left. \lambda 
	\right] \mu a b} = 0 \ .\label{comcurve}
\end{eqnarray}
Note also that the combination of Equations \pref{nienhuis} and \pref{delcom}
leads to an algebraic 
constraint on the torsion (Delduc, Kalitzin and Sokatchev 1990):
\begin{equation}
H^{\mu}{}_{\nu\rho} 
	- H^{\mu}{}_{\lambda\tau}J^{\lambda}{}_{\tau}J^{\tau}{}_{\rho}
	+ H^{\tau}{}_{\nu\lambda}J^{\mu}{}_{\tau}J^{\lambda}{}_{\rho}
	+ H^{\tau}{}_{\lambda\rho}J^{\mu}{}_{\tau}J^{\lambda}{}_{\nu}
	= 0\ .\label{torcon}
\end{equation}
Any two of Equations \pref{nienhuis}, \pref{delcom}, \pref{torcon}
are independent. 
One may also rewrite Equations  
\pref{jeqi} - \pref{torcon} as the vanishing of
various components of the connection, torsion, and curvature
in complex coordinates (Hull and Witten 1985,
Bonneau and Valent 1994).

\section{Construction of SUSY and $U(1)_{L}$ currents}

We will describe the left-moving $N=1$ supersymmetry and $U(1)$
via their current algebra.  Although we will work only to
$\OO(\apr^{0})$, the supersymmetry is realized
quantum mechanically on the internal fermions and the algebra
only closes to this order 
after including worldsheet loop effects in the commutators.

We begin by building the supersymmetry current out of the most general
terms which have dimension $(0,3/2)$ and the $U(1)$ current out of 
terms which have dimension
$(0,1)$.  We then ask that the algebra close properly and that the
variation of the action under the classical transformations
vanish up to the divergence of the currents, or, integrating by parts,
up to terms of the form
\begin{equation}
	\p_{-}\epsilon J_{+} + \p_{+}\epsilon J_{-}\ ,
\end{equation}
where $\epsilon$ parameterizes the variation, 
$\delta\xi=\{\xi,\epsilon J\}$.
Since we want left-moving currents we also demand that $J_{+}$ vanish.

To order $\apr^{0}$ the equal-time commutators are usually just the
tree-level Poisson or Dirac brackets of the currents.
However, as we have discussed above,
supersymmetry is realized nonlinearly and quantum mechanically
on the internal fermions.
It is easy to see that the operator product of the supersymmetry
current \pref{fermsusy} with itself is:
\begin{equation}
G(w_{-}) G(z_{-}) \sim i \left(\apr\right)^{2} \frac{d}{12} 
	\frac{1}{(w_{-} - z_{-})^{3}}
	- \apr \frac{1}{(w_{-}-z_{-})} \eta_{ab}\lambda^{a}
		\p_{-}\lambda^{b}\ ,
\end{equation}
where we have explicitly included the $\apr$ factors coming from fermion
loops.  This is the correct operator product for the left-moving SUSY charge
of a system with central charge $\hat{c} = d/3$ within a multiplicative 
factor of $\apr$ (for
a discussion of this factor in path integral language see 
(Gates, Howe and Hull 1989)).  To get operator product
coefficients of the superconformal algebra we
must rescale $G$:
\begin{equation}
G = \frac{1}{3\sqrt{\apr C_{A}}} f_{abc}\lambda^{a}\lambda^{b}\lambda^{c}\ .
\label{scalesusy}
\end{equation}
One may also see this scaling by starting with the 
left-moving supersymmetry current
of a scalar field plus its fermionic superpartner:
\begin{equation}
G_{L} = \p_{-}\phi\lambda^{1}\ .
\label{clasusy}
\end{equation}
Now fermionize
the scalar:
\begin{equation}
\sqrt{\frac{\apr}{2}} \p_{-} \phi = : \lambda^{2}\lambda^{3} :\ .
\end{equation}
The coefficient on the left-hand side of this equation, 
including the power of $\apr$, can be fixed
by matching the two-point function of each side.
Substituting this into \pref{clasusy}
will give us a supersymmetry current like \pref{scalesusy}.
Because of this $\apr$ dependence, we must be a bit careful in counting 
orders of $\apr$ in
our computation; for example, one loop terms in operator products of
the above supersymmetry charge will enter at order ${\apr}^{0}$.

In previous discussions of $\sigma$-models with $(2,1)$ supersymmetry
(Hull 1986b; Dine and Seiberg 1986; Braden 1987)
closure of the algebra was imposed at tree level.
This forces the
gauge fields to be flat and the internal fermions to be non-propagating.
However, in flat space there are physical vertex operators corresponding to
gauge field fluctuations; nothing prevents us from
condensing these operators 
on the worldsheet as long as their expectation values
satisfy the $\beta$-function equations.
The latter three authors listed above argued
that because supersymmetry pairs bosons and fermions, the internal fermions
must be trivial since they are not paired with bosonic fields.
However, in the full quantum treatment
we may have a representation of supersymmetry
on the Fock space of the system which does not close
on the one-particle states.

With these arguments in mind, we also expect that the part of the
left-moving $U(1)$ composed of left-moving fermions will have a piece
quadratic in fermions and scaling as $1/\sqrt{\apr}$,
$\Sigma_{ab}\lambda^{a}\lambda^{b}$.  This is especially necessary
for the $U(1)$ which leads to membrane constructions
(Kutasov and Martinec 1996; Kutasov, Martinec and O'Loughlin 1996).
For these constructions the part of the $U(1)$ living in
the target space is
timelike; for example, if the current in flat space looks like:
\begin{equation}
J_{-} = v_{\mu} \p_{-}\phi^{\mu} + \Sigma_{ab}\lambda^{a}\lambda^{b}\ ,
\end{equation}
and $v^{2} < 0$, then the anomaly in the
operator product of this with current can only vanish by setting
$\apr \Sigma_{ab}\Sigma^{ab} + v_{\mu}v^{\mu} = 0$, where the first
term is a one-loop term.  Thus, $\Sigma$ must 
generally have a piece scaling as $1/\sqrt{\apr}$.

\subsection{Review of $(2,0)$ SUSY charges}

The right-moving currents for $N=2$ SUSY can be written down most easily
as $(1,0)$ superfields (see for example Hull and Spence 1990):
\begin{eqnarray}
\GG^{(1)}_{+} & = & g_{\mu\nu}\left(\Phi\right)D_{+}\Phi^{\mu}\p_{+}\Phi^{\nu}
	- \frac{i}{6} H_{\mu\nu\rho}\left(\Phi\right)
	  D_{+}\Phi^{\mu}D_{+}\Phi^{\nu}D_{+}\Phi^{\rho} \\
\UU_{+} & = & \frac{i}{2} J_{\mu\nu}\left(\Phi\right) 
		D_{+}\Phi^{\mu}D_{+}\Phi^{\nu}\ .
\label{rcurrsf}
\end{eqnarray}
Expanding the superfields in components we find (see Appendix A for
notation):
\begin{eqnarray}
\GG_{+}^{(1)} & = & g_{\mu\nu}\psi^{\mu}\p_{+}\phi^{\nu}
	- \frac{i}{6} H_{\mu\nu\rho}\psi^{\mu}\psi^{\nu}\psi^{\rho}
		\nonumber \\
	&& \ \ \ + i \theta_{+} \left\{ g_{\mu\nu}\p_{+}\phi^{\mu}
		\p_{+}\phi^{\nu}
	+ i g_{\mu\nu}\psi^{\mu}\left(\p_{+}\psi^{\nu}
	+ {\Gamma_{(+)}}^{\nu}_{\lambda\rho}\p_{+}\phi^{\lambda}\psi^{\rho}
		\right)\right\} \nonumber \\
	& = & G^{(1)}_{+} + i \theta T_{++} \label{rightg1}\\
\UU_{R} & = & \frac{i}{2} J_{\mu\nu}\psi^{\mu}\psi^{\nu}
	+ \theta \left\{ J_{\mu\nu}\psi^{\mu}\p_{+}\phi^{\nu}
		+ \frac{i}{2} \p_{\rho}J_{\mu\nu}
		\psi^{\mu}\psi^{\nu}\psi^{\rho} \right\} \nonumber \\
	& = & J_{+} + \theta G^{(2)}_{+}\ .
\end{eqnarray}
Here $G^{(1)}_{+}$, $G^{(2)}_{+}$ are the two right-moving 
supersymmetry currents,
$T_{++}$ is the right-moving stress-tensor, and $J_{+}$ is the
$U(1)$ part of the right-moving superconformal algebra.
It is easy to see, using the fundamental
Dirac brackets described in Appendix B, that the above currents generate
the transformations listed in Equations \pref{n1rphi}-\pref{rcstruc}.

\subsection{The left-moving currents}

The most general dimension $(0,3/2)$ operator we can write down is:
\begin{equation}
G_{-} = e_{\mu a}\p_{-}\phi^{\mu}\lambda^{a} 
		- i f_{abc}\lambda^{a}\lambda^{b}\lambda^{c}\ .
\label{origlsusy}
\end{equation}
The variations of the component fields induced by
Dirac brackets with this current are:
\begin{eqnarray}
\delta \phi^{\mu} & = & \epsilon\, e^{\mu}{}_{a}\lambda^{a} \label{lsphi}\\
\delta \psi^{\mu} & = & \epsilon\, e^{\rho}{}_{a} 
	{\Gamma_{(+)}}^{\mu}_{\rho\alpha} \psi^{\alpha}\lambda^{a} 
		\label{lspsi}\\
\delta \lambda^{a} & = & i \epsilon e_{\mu}{}^{a}\p_{-}\phi^{\mu}
		+ \epsilon \left(3 f^{a}{}_{bc} - 
	e^{\rho}{}_{b}{\hat{A}}_{\rho}{}^{a}{}_{c}\right)
		\lambda^{a}\lambda^{b} \nonumber \\
	&\equiv& i \epsilon e_{\mu}{}^{a}\p_{-}\phi^{\mu} +
		\epsilon B^{a}{}_{bc} \lambda^{b}\lambda^{c}\label{lslam}
\end{eqnarray}

The object $e_{\mu a}$ maps the 4-dimensional
tangent bundle of the target worldvolume 
into the 28-dimensional vector bundle $\VV$.
We may also form the projector (Braden 1987)
\begin{equation}
\PP^{a}{}_{b} = \delta^{a}{}_{b} - e_{\mu}{}^{a} e^{\mu}{}_{b}\ .
\end{equation}
This projects indices onto what we will call the normal part of
$\VV$ while $(1-\PP)$ projects indices onto the tangent part
of $\VV$: if $e$ has maximal rank (which should be true for
generic points in the target worldvolume), the normal part
of the bundle will have
$24$-dimensional fibers and the tangent part will have $4$-dimensional
fibers.
In flat space, as we have discussed,
the fermions will lie in the adjoint representation
of some Lie algebra and the normal part of $f^{abc}$ will be the 
structure constants of the algebra.  We expect that this structure will
persist fiberwise in the presence of nontrivial $g_{\mu\nu}$,
$\hat{A}$, and $e$.

The most general dimension $(0,1)$ operator is:
\begin{equation}
J_{-} = v_{\mu} \p_{-}\phi^{\mu} + 
	\Sigma_{ab}\lambda^{a}\lambda^{b}\ ,
\label{lsusycur}
\end{equation}
where all the coefficients depend on $\phi$ as usual.
The variations of the $\sigma$-model variables under this current are:
\begin{eqnarray}
\delta\phi^{\mu} &=& \epsilon\, v^{\mu} \label{ljphi} \\
\delta\psi^{\mu} &=& - \epsilon\, 
	{\Gamma_{(-)}}^{\mu}_{\nu\rho}v^{\rho}\psi^{\nu} 
\label{ljpsi} \\
\delta\lambda^{a} &=& -\epsilon\, 
	{\hat{A}_{\rho}}{}^{a}{}_{b}v^{\rho}\lambda^{b}
	- 2 i \Sigma^{a}{}_{b}\lambda^{b}\ .
\label{ljlam}
\end{eqnarray}
$v$ and $\Sigma$ will be restricted by demanding chirality,
invariance of the action under the above variations, and closure
of the left-moving algebra.

\section{Imposing $N=(0,1)$ SUSY}

As with $(2,2)$ (Alvarez-Gaum\'{e} and Freedman 1981) and $(2,0)$ 
(Hull and Witten 1986) supersymmetry in 2d $\sigma$-models,
$(2,1)$ supersymmetry places constraints on the geometry of the
target space fields.
We find our constraints in the usual fashion,
by demanding the classical invariance of
the action under the variations \pref{lsphi}-\pref{lslam},
and closure of the supersymmetry algebra.
The constraints arising from the right-moving supersymmetries
have been reviewed above.
The constraints
on fields coupled to the tangent fermions,
were found by Braden (1987); we will rederive his
results in the course of our analysis.

In the variation of the action we can clearly separate the
terms into pieces of different order in the left- and right-moving
fermions and their derivatives.  The bosonic part of the
action will give us all the terms linear in $\lambda$.  After several
integrations by parts and the use of the equations of motion for
$\phi$, we find that the term proportional to $\lambda$ is:
\begin{eqnarray}
\lefteqn{\delta S = 
	\int d^{2}\sigma \left\{ \p_{+} \epsilon\, e_{\mu a} \lambda^{a}
	\p_{-}\phi^{\mu} \right.} \nonumber \\
&&\ \ + \epsilon\! \left. \left[\p_{\mu}e_{\nu a} - 
		\hat{A}{}_{\mu}{}^{b}{}_{a} e_{b\nu}
	- {\Gamma_{(+)}}{}^{\alpha}_{\nu\mu}e_{\alpha a}\right]
	\lambda^{a}\p_{+}\phi^{\mu}\p_{-}\phi^{\nu} \right\}\ .
\label{linvar}
\end{eqnarray}
The first term gives us the first piece of the supersymmetry current
\pref{lsusycur} as required by Noether's theorem.  The vanishing of
the second term
was interpreted by Braden (1987) as an equality between the
spin and (tangent bundle) gauge connections up to a (sort of) 
gauge transformation.
Another interpretation is that $e$ is covariantly 
constant to $\OO(\apr^{0})$:
\begin{equation}
\hat{D}_{(-)\mu} e^{\rho}{}_{a} = 0\ ,
\end{equation}
which implies that
\begin{equation}
\hat{D}_{\lambda} \PP^{a}{}_{b} = 0
\end{equation}
at $\OO(\apr^{0})$.
The terms proportional to $\psi\psi\lambda$ are:
\begin{equation}
\int i \epsilon \left[ R_{(+)\mu\nu\lambda\rho} e^{\lambda}{}_{a}
	- F_{\mu\nu ab} e^{b}{}_{\rho}\right]\p_{-}\phi^{\rho}
		\lambda^{a}\psi^{\mu}\psi^{\nu}\ .
\label{reqf1}
\end{equation}
This sets the tangent part of $F_{ab}$ equal to $R_{(+)}$,
and also implies that $F_{ab}$ splits entirely into tangent and
normal pieces.
The term cubic in $\lambda$ is:
\begin{eqnarray}
\lefteqn{T_{3\lambda} = \int d^{2}\sigma\left\{
	-i \p_{+}\epsilon f_{abc}\lambda^{a}\lambda^{b}\lambda^{c}
		\right.} \nonumber \\
	&&\left.+ i\epsilon\left(\frac{1}{2} e^{\alpha}{}_{a}
		F_{\alpha\rho}{}_{bc}
	- \hat{D}_{\rho}f_{abc}\right)\p_{+}\phi^{\rho}
	\lambda^{a}\lambda^{b}\lambda^{c}\right\}\ .
\label{threelam}
\end{eqnarray}
The first term multiplying $\p_{+} \epsilon$ combines with the first term
of the integrand in Equation \pref{linvar} to form the supersymmetry current, 
as required by Noether's theorem.
The second term will be discussed below.  Finally, there is a term
cubic in $\lambda$ and quadratic in $\psi$, arising
from the variation of the $F\psi\psi\lambda\lambda$ term
in the action:
\begin{equation}
T_{2\psi 3\lambda} = \int d^{2}\sigma \left\{\left(
	e^{\rho}{}_{a}\hat{D}^{(+)}_{\rho}F_{\mu\nu bc}
	+ 6 F_{\mu\nu k a}f^{k}{}_{bc}\right)
	\psi^{\mu}\psi^{\mu}\lambda^{a}\lambda^{b}\lambda^{c}\right\}\ .
\label{2p3l}
\end{equation}

Next, we wish to examine the constraints arising from closure of
the supersymmetry algebra:
\begin{eqnarray}
\{G_{-}(\sigma),G_{-}(\sigma')\} &=& T_{- -}(\sigma')\ \ 
	\delta (\sigma-\sigma') \\
\{G_{-}(\sigma),G_{+}^{(1)}(\sigma')\} &=& 0 \\
\{G_{-}(\sigma),J_{+}(\sigma')\} &=& 0 \\
\{G_{-}(\sigma),G_{+}^{(2)}(\sigma')\} &=& 0\ .
\end{eqnarray}
Note that the last equation follows from the first three by the Jacobi
identity.  We will be interested in the $\OO(\apr^{0})$ part of
these commutators.  As discussed above the first of these 
commutators will have two pieces at order $\OO(\apr^{0})$:
one will arise from the classical Dirac bracket and the other will
arise from the one-loop contribution of the commutator of
the cubic part of the supersymmetry current with itself.
A long and tedious calculation, using the results above,
reveals the classical Dirac brackets of the
left-moving SUSY currents to be:
\begin{eqnarray}
\lefteqn{\left\{ G_{-}(\sigma),G_{-}(\sigma')\right\}=
	\delta (\sigma-\sigma') \times}
	\nonumber \\
&&\ \ \ \left\{ -i e_{\mu}{}^{a}e_{\nu a}\p_{-}\phi^{\mu}\p_{-}\phi^{\nu}
	+ e_{\rho a}e^{\rho}{}_{b}\lambda^{a}\left(
	\p_{-}\lambda^{b} + \hat{A}_{\mu}{}^{b}{}_{c}\p_{-}\phi^{\mu}
		\lambda^{c}\right) \right.\nonumber \\
&&\ \ \ - \left[ \left( 6 f_{abc}e_{\rho}{}^{c}+H_{\alpha\beta\rho}
	e^{\alpha}{}_{a}e^{\beta}{}_{b}\right) \right. \nonumber \\
&&\ \ \ \left. + 2e^{\mu}{}_{a}\hat{D}_{(-)\mu}e_{\rho b}
			- e^{\mu}{}_{a}\hat{D}_{(-)\rho}e_{\mu b}
		\right]
		\p_{-}\phi^{\mu}\lambda^{a}
		\lambda^{b} \nonumber \\
&&\ \ \ - e^{\mu}{}_{a}\hat{D}_{(-)\rho}e_{\mu b}\p_{+}
	\phi^{\rho}\lambda^{a}\lambda^{b} \nonumber \\
&&\ \ \ + \frac{i}{2}\left(R_{(+)\lambda\rho\alpha\beta}e^{\alpha}{}_{a}
		e^{\beta}{}_{b}
	- e_{\mu a}e^{\mu}{}_{k} F_{\lambda\rho}{}^{k}{}_{b}
	\right)\lambda^{a}\lambda^{b}\psi^{\lambda}\psi^{\rho}
	\nonumber\\
&&\ \ \ \left. + i\left(\frac{1}{2}e^{\alpha}{}_{a}e^{\beta}{}_{b}
	F_{\alpha\beta cd} +
	2 e^{\alpha}{}_{a}\hat{D}_{\alpha}f_{bcd} +
	9 f^{k}{}_{ab}f_{kbc}\right)
		\lambda^{a}\lambda^{b}\lambda^{c}\lambda^{d}\right\}\ .
\label{glcomm}
\end{eqnarray}
Note that we had to use the equations of motion for $\lambda$ in order
to get terms with $\tau$ derivatives of $\lambda$.
The first line of \pref{glcomm}
is clearly the stress tensor of the bosons and of the
fermions tangent to the target worldvolume, if
\begin{equation}
e_{\mu}{}^{a}e_{\nu a} = g_{\mu\nu}\ .
\label{esquared}
\end{equation}
We have kept the 
$\hat{D}_{(-)}e$ terms in the fourth and fifth line.  
They vanish to this order, but we will want
to discuss order $\OO(\sqrt{\apr})$ corrections below.
To order $\OO(\apr^{0})$ the vanishing of the
third line of \pref{glcomm} 
requires $f$ to split into a completely normal piece $f^{\perp}$
and a completely tangent piece:
\begin{equation}
f^{\|}{}_{abc} = - \frac{1}{6} e^{\alpha}{}_{a}
	e^{\beta}{}_{b}e^{\gamma}{}_{c}H_{\alpha\beta\gamma}\ .
\label{feqh}
\end{equation}
As we will see shortly, corrections of this splitting at
$\OO(\sqrt{\apr})$ mix into the last line of Equation \pref{glcomm}.  
The vanishing of the sixth line follows from
the vanishing of \pref{reqf1}.  The final line has a piece of order
$1/\apr$.  If we were to think of $f$ as some
structure constants on the normal part of the fibers of the bundle $V$, 
then the vanishing of this term, 
\begin{equation}
f^{(\perp)}{}_{k\left[ ab\right.} f^{(\perp)k}{}_{\left. cd\right]} = 0\ ,
\label{jacobi}
\end{equation}
enforces the Jacobi identity on the structure constants.  Note
that if $f$ receives an $\OO(\sqrt{\apr})$ correction $f^{(1)}$, then
the $ff$ term will have an additional $\OO(\apr^{0})$ term
coming from $f^{\perp}f^{(1)}$.  At order $\OO(1/\sqrt{\apr})$,
\begin{equation}
\hat{D}_{\lambda}f^{\perp}{}_{abc} = 0\ .
\end{equation}
This equation will have $\OO(\apr^{0})$ 
corrections.

\figloc1{The one-loop $\OO(\apr^{0})$ contribution to the commutator
$\{G_{-},G_{-}\}$.  The single straight
lines denote $\lambda$ propagators; the triple lines denote
background field insertions.  Crossed circles denote vertices arising
from operator insertions.}

We must also examine the one loop contribution 
to the commutator of $f^{(\perp)}{}_{abc}\lambda^{a}\lambda^{b}\lambda^{c}$ 
with itself.  This can be calculated
by expanding the operators and the action in Riemannian normal coordinates
(Alvarez-Gaum\'{e}, Freedman
and Mukhi 1981; Sen 1985; Banks, Nemeschansky and Sen 1986).  We can
get away with calculating a subset
of the resulting terms since the normal coordinate expansion is
covariant with respect to both the target space
gauge and coordinate indices.
Terms coming from expansions
of $f$ in normal coordinates will
involve derivatives of $f$ which will be covariantized; as argued above,
such terms will vanish at $\OO(\apr^{0})$ (although not
necessarily at $\OO(\sqrt{\apr})$).  Terms
in the commutator involving the gauge and spin connection
will either covariantize derivatives
or form combinations and covariant derivatives
of the appropriate curvature tensors.  

In general this argument is too naive.  The $\sigma$-model
anomaly (Moore and Nelson 1984,1985; Hull and Witten 1985) 
spoils invariance with respect to
local Lorentz and gauge transformations of the background fields.
This lack of gauge invariance can be absorbed by
an anomalous variation of $b_{\mu\nu}$ and
a redefinition of the antisymmetric tensor field
strength (Callan \etal\ 1985; Hull and Witten 1985; Sen 1986;
Hull and Townsend 1986).
At any rate, the $\sigma$-model anomaly will not show up
to the order we are concerned with.  Another potential
problem arises if we expand the action around a background
field configuration which does not satisfy the equations
of motion; in this case, the normal
coordinate expansion of the action will include noncovariant
terms proportional to the classical equations of motion
(Alvarez-Gaum\'{e}, Freedman and Mukhi 1981).  This leads to
noncovariant divergences in the action which are removed
by wavefunction
renormalization.  Such terms should be included when renormalizing
the theory, but once this is done we can calculate the renormalized
Green's
functions by expanding the action around solutions to the
classical equations of motion.

\figloc2{The remaining one-loop $\OO(\apr^{0})$ diagrams, after Figure 1.
The open circles denote vertices arising from the interaction part of the
Lagrangian.}

The relevant one-loop Feynman diagrams contributing to the commutator are
shown in Figures 1 and 2.
The diagram in Figure 1
will give us a coefficient of $1/(x_{-}-y_{-})^{2}$ times a bilocal operator:
expanding the operator around $y$ gives us a term multiplying
$1/(x_{-}-y_{-})$ which will contain derivatives of $f$ and $\lambda$ which
will become covariantized.  The diagrams in
Figure 2 cancel each other due to the Jacobi identity \pref{jacobi}.
The result for the one-loop $\OO(\apr^{0})$ operator product is:
\begin{eqnarray}
\lefteqn{G_{-}(x_{-}) G_{-}(y_{-}) = -\frac{18}{4}
	\frac{\apr^{2}}{\left(x_{-}-y_{-}\right)^{2}}
	f^{(\perp)}{}_{A}{}^{CD}f^{(\perp)}{}_{BCD}\lambda^{A}\lambda^{B}
	(y_{-})} \nonumber \\
	&&\ \ \ +\frac{18}{4}\frac{\apr}{\left(x_{-}-y_{-}\right)}
	f^{(\perp)}{}_{A}{}^{CD}f^{(\perp)}{}_{BCD}
	\lambda^{A}\left(\p_{-}\lambda^{B}
		+ \omega_{\mu}{}^{B}{}_{K}\p_{-}\phi^{\mu}\lambda^{K}\right)\ .
\end{eqnarray}
The last term in the operator product is equal to 
$iT^{(\perp)}/2(x_{-}-y_{-})$, as required for closure, provided that
\begin{equation}
f^{(\perp)}{}_{ACD}f^{(\perp)}{}_{B}{}^{CD} = 
	- \frac{1}{9\apr} \PP^{K}{}_{A}\PP^{L}{}_{B}\eta_{KL} \ ,
\label{ftog}
\end{equation}
This equation and Equation \pref{jacobi} indicates that the
coefficients $f^{(\perp)}$
are proportional to structure constants of a Lie algebra.
Equation \pref{ftog} also insures that the $1/(x_{-}-y_{-})^{2}$ term
vanishes, and gives the $1/(x_{-}-y_{-})^{3}$ term expected for
24 free fermions.  The remaining singular part of the
operator product can be converted
to an expression for the equal-time commutator in the usual
fashion (see Appendix C).


If Equation \pref{ftog}  satisfied we may add to
Equation \pref{glcomm} the term
\begin{equation}
\PP^{k}{}_{a}\PP_{kb}\lambda^{a}\left(\p_{-}\lambda^{b}
	+ \hat{A}_{\mu}{}^{b}{}_{c}\lambda^{c}\right)
\end{equation}
at order $\OO(\apr)$.  Adding this term to the second line of
Equation \pref{glcomm} should give us $T_{--}$ as required for
closure.

Another long calculation reveals that to
$\OO(\apr^{0})$
\begin{equation}
\left\{G_{-}(\sigma),G_{+}(\sigma')\right\}
= i \left(\frac{1}{2}e^{\alpha}{}_{a}F_{\alpha\rho b c}
	- \hat{D}_{\rho}f_{abc}\right)
	\psi^{\rho}\lambda^{a}\lambda^{b}\lambda^{c}\ .
\label{glrcomm}
\end{equation}
The vanishing of this term follows from the vanishing of \pref{threelam}.
Note that since there are no terms with negative powers of
$\sqrt{\apr}$ appearing in $G_{+}$ all the terms in
\pref{glrcomm} will arise at tree level.

Now we can discuss the solution to the constraints we have derived.
The operator product contains no piece of
$G_{ab}\lambda^{a}\p_{-}\lambda^{b}$ where $G$ has both
tangent and normal indices: in other words, the metric
is block diagonal with respect to the splitting of $\VV$
defined by $\PP$.
In addition, the vanishing of
\pref{reqf1} indicates that $F$ is also block diagonal
in its gauge indices at $\OO(\apr^{0})$.  If we split $F$
into $F^{(\|)}$ and $F^{(\perp)}$, and if we let
$f=f^{(\perp)} - iH/6$, we find that using the Bianchi identities in the
appropriate manner, terms involving $F^{\|}$ and $H$ drop out
of the the last line of Equations \pref{glcomm}, \pref{threelam}, and
\pref{2p3l}. This leaves the following constraints:
\begin{eqnarray}
&&\left\{\frac{1}{2} F^{(\perp)}{}_{\mu\rho bc }
	e^{\mu}{}_{a} 
	- \hat{D}_{\rho}f^{(\perp)}{}_{abc} = 0
	\right\}\lambda^{a}\lambda^{b}\lambda^{c} \label{cons1}\\
&&\left\{\frac{1}{2}e^{\alpha}{}_{a}e^{\beta}{}_{b}
		F^{(\perp)}{}_{\alpha\beta cd}
	+ 2e^{\alpha}{}_{a}\hat{D}_{\alpha}f^{(\perp)}
		{}_{bcd}
	+ 9f^{(\perp)}_{k ab}f^{(\perp)k}{}_{cd}\right\}
	\lambda^{a}\lambda^{b}\lambda^{c}\lambda^{d} 
		= 0 \label{cons2} \\
&&\left\{\hat{D}_{(+)\rho}F^{(\perp)}_{\mu\nu bc}
	e^{\rho}{}_{a}
	+ 6F^{(\perp)}_{\mu\nu ka} f^{(\perp)k}
		{}_{bc}\right\}
	\lambda^{a}\lambda^{b}\lambda^{c} = 0\ . \label{cons3}
\end{eqnarray}
It seems at first glance impossible to solve these constraints for
nontrivial $F^{(\perp)}$. 
However,
order $\OO(\sqrt{\apr})$ terms in the supersymmetry current combine with
order $\OO(1/\sqrt{\apr})$ terms in the above equations
and cancel off the nontrivial
field strengths.  Closure to order
$\OO(\sqrt{\apr})$ requires that one-loop terms in
the commutator cancel the classical $\OO(\sqrt{\apr})$
terms: we will calculate one of these terms below.

Let us assume that $f$ has a term $f^{(1)}$ scaling as $\sqrt{\apr}$
with one normal index and two tangent indices.  If we substitute Equation
\pref{cons2} into Equation \pref{cons1}, and use the version of
Equation \pref{ftog} with curved indices,
\begin{equation}
f^{(\perp)}{}_{akl}f^{(\perp)}{}_{b}{}^{kl} = - \frac{1}{9} G_{ab}\ ,
\label{cartanmetric}
\end{equation}
we find that
\begin{equation}
f^{(1)}_{abk} \PP^{k}{}_{c} = - \frac{1}{4}\apr
	e^{\alpha}{}_{a}e^{\beta}{}_{b}
		F_{\alpha\beta kl} f^{(\perp)}{}_{c}{}^{kl}\ .
\label{fcorr}
\end{equation}

We may solve Equation \pref{cons3} by adding an order $\OO(\sqrt{\apr})$ 
piece $\hat{A}^{(1)}$ to the gauge connection $\hat{A}$:
\begin{equation}
\hat{A}^{(1)}{}_{\mu}{}^{a}{}_{b}e^{b}{}_{\rho} = \frac{3}{2} \apr
	F^{(0)}{}_{\rho\mu kl} f^{(\perp) akl}\ ,
\label{conncorr}
\end{equation}
where $F^{(0)}$ is the curvature for the uncorrected connection
$\hat{A}-\hat{A}^{(1)}$.  If $\hat{D}^{(0)}$ is the gauge-covariant
derivative with respect to the uncorrected connection, then the order 
$\OO(\sqrt{\apr})$ correction to the curvature is:
\begin{equation}
F_{\mu\nu}{}^{a}{}_{b} = F^{(0)}{}_{\mu\nu}{}^{a}{}_{b}
	+ \hat{D}^{(0)}{}_{\mu}\hat{A}^{(1)}{}_{\nu}{}^{a}{}_{b}
	- \hat{D}^{(0)}{}_{\nu}\hat{A}^{(1)}{}_{\mu}{}^{a}{}_{b}
	+ \OO(\apr)\ .
\end{equation}
Note that this will imply that $F$ has mixed tangent and normal
indices at order $\sqrt{\apr}$.
After some manipulation, Equation \pref{cons3} becomes:
\begin{equation}
\hat{D}^{(0)}_{\left[\mu\right. }
	F^{(0)}{}_{\left. \nu\rho\right]kl}f^{(\perp)}{}_{a}{}^{kl}
	= 0\ ,
\end{equation}
which is the Bianchi identity for $F$ to $\OO(\apr^{0})$.

Since $f^{(\perp)}$ has entirely normal indices at
$\OO(\apr^{0})$, we may write
\begin{equation}
f^{(\perp)}{}_{abc} = \PP^{k}{}_{a}f^{(\perp)}{}_{kbc}\ .
\end{equation}
Applying $\hat{D}$ to both sides, we find that
\begin{equation} 
\left(1 - \PP\right)^{k}{}_{a}\hat{D}_{\rho}f^{(\perp)}{}_{kbc}
	= \left(\hat{D}_{\rho}\PP^{k}{}_{a}\right) f^{(\perp)}{}_{kbc}\ .
\label{covp1}
\end{equation}
Combined with Equation \pref{cons2}, this means that:
\begin{equation}
\hat{D}_{\mu} e^{a}{}_{\rho} = -\apr\frac{3}{2}
	F^{(0)}{}_{\mu\rho kl}f^{(\perp)a kl}\ .
\label{cove}
\end{equation}

We have found all of the constraints on the target space geometry and
the left-moving supersymmetry current
up to $\OO(\apr^{0})$.  Let us
summarize these results.  The left-moving supersymmetry
identifies a ``tangent'' subbundle of $\VV$ with the
tangent bundle of the target space; it also requires that the
fibers of the orthogonal bundle are acted on by some Lie algebra in the
adjoint representation.  The
one-form $e_{\mu}{}^{a}$ identifies the tangent subbundle of
$\VV$.  The geometric structures of the target space and the
tangent part of $\VV$ are identified by
the vanishing of \pref{reqf1}
\begin{equation}
R_{(+)\mu\nu\lambda\rho}=F_{\mu\nu ab}e_{\lambda}{}^{a}
	e_{\rho}{}^{b}\ , \label{reqf2}
\end{equation}
and by the identification of the tangent part of $f$ with the torsion
in Equation \pref{feqh}.  Note that with this identification the
tangent part of $G_{-}$ is the left-moving analog of the supersymmetry
current in Equation \pref{rightg1}.  Equation \pref{cove} constrains
the mapping $e_{\mu}{}^{a}$.
Note that if the right hand side of Equation
\pref{cove} is proportional to gauge and Lorentz covariant terms,
then by dimensional analysis these terms must scale at least as $\sqrt{\apr}$,
which is the only length scale available.  The structure on the
fibers of the normal subbundle of $\VV$ is encoded in $f^{(\perp)}$.
Equations \pref{jacobi} and \pref{cartanmetric} force $f^{(\perp)}$
to be, fiber by fiber, structure constants of some Lie group.
The rotation of this group structure as we move around the target
space is constrained by Equation \pref{cons1};
this equation can be derived from Equation \pref{cove}.
The remaining constraints consist of $\OO(\sqrt{\apr})$ modifications
of $\hat{A}$ and $f$; these modifications arise from nontrivial
transverse gauge fields.   The above constraints in the presence
of transverse gauge curvature are the major new results of this
section.

\figloc3{A one-loop $\OO ( \apr^{1/2} )$ contribution to the
commutator $\left\{ G_{-},G_{-}\right\}$. The 
wavy line denotes a boson propagator. }

Since the constraints we have found
require adding order $\OO(\sqrt{\apr})$ terms to
the action and to the supersymmetry current, the next step would be to
check closure of the supersymmetry algebra at order $\OO(\sqrt{\apr})$.
We will not pursue this very far; however, to reassure ourselves that
our solutions make sense at higher order, let us discuss
one-loop corrections to the terms in Equation \pref{glcomm} which
are quadratic in $\lambda$.  The correction to $\hat{A}$ will
show up in the first line of Equation \pref{glcomm}
as an order $\OO(\sqrt{\apr})$ piece of 
$T_{--}$, but this is due to the addition of $\hat{A}^{(1)}$ to the
connection.  In the second line $f^{(1)}$ and the order $\sqrt{\apr}$
corrections to $\hat{D}_{(-)}e$ combine to form the order $\sqrt{\apr}$
term in the second line of Equation \pref{glcomm}:
\begin{equation}
\apr \frac{3}{2}e^{\mu}{}_{a}F^{(0)}{}_{\mu\rho kl}
	f^{(\perp)}{}_{b}{}^{kl}\p_{-}\phi^{\rho}\lambda^{a}\lambda^{b}\ .
\end{equation}
Closure requires that this be cancelled by one loop contributions to
\pref{glcomm} at the appropriate order: such contributions 
will come from the
commutator of $e\p_{-}\phi\lambda$ with $f\lambda\lambda\lambda$.
The diagram leading to the one-loop term proportional to
\begin{equation}
\apr e^{\mu}{}_{a}\p_{\rho}\hat{A}_{\mu kl} f_{b}{}^{kl}
	\p_{-}\phi^{\rho}\lambda^{a}\lambda^{b}
\label{diag1}
\end{equation}
is shown in Figure 3.  If we evaluate this diagram using the conventions
stated in Appendix C, we find that the coefficient is $3/2$.  Note that the
boson propagator in this diagram corresponds to the Green's
function
\begin{equation}
\langle \p_{+}\phi(\sigma)\p_{-}\phi(\sigma')\rangle = 2\pi i 
	\delta(\sigma-\sigma')\ .
\end{equation}
Since
the normal coordinate expansion is manifestly gauge and coordinate
covariant, all terms in the commutator at this order contain $\hat{A}$
either in covariant derivatives or in curvature tensors
(although for higher orders we should keep in mind the $\sigma$-model
anomaly).  Thus, so long
as we know the correct coefficient in front of the term \pref{diag1}
we know the coefficient multiplying the term proportional to
gauge curvature, of which \pref{diag1} is a piece.  In this case, then, the
one loop order $\OO(\sqrt{\apr})$ contribution to Equation
\pref{glcomm} proportional to the gauge curvature is:
\begin{equation}
\apr \frac{3}{2}e^{\mu}{}_{a}F^{(0)}{}_{\rho\mu kl}
	f^{(\perp)}{}_{b}{}^{kl}\p_{-}\phi^{\rho}\lambda^{a}\lambda^{b}\ ,
\end{equation}
which cancels the contribution from $f^{(1)}$ and $\hat{D}e$.  
We will leave the calculations of further one- and higher-loop contributions
to the current commutators for future work.  

We would like to note
that although the $1/\sqrt{\apr}$ scaling complicates the $\apr$
expansion of the commutators it does not invalidate this
expansion.  The only negative powers of $\sqrt{\apr}$ arise
from the currents themselves.  $\sigma$-model counterterms will
always be multiplied by positive powers of $\apr$ with respect
to the bare Lagrangian; the divergences of the Green's functions
may scale with large negative powers of $\apr$ if there are enough
supersymmetry current insertions, but this is due
to the scaling of external sources.  One- and higher-loop
divergences will always be higher order than the
tree-level Green's function and will be removed by counterterms
scaling with positive powers of $\apr$.

\section{Imposing the $U(1)_{L}$ symmetry}

The restrictions on the geometry of the $(2,1)$ $\sigma$-model
due to the $U(1)$ symmetry are derived in the same way as
in the previous section.  We require that the action be invariant;
that the $U(1)$ current algebra contain no central term;
and that the current be chiral:
\begin{equation}
\p_{+}J_{-} = 0 \ .
\end{equation}
The conditions arising from invariance of the action and from chirality have
been derived for the bosonic $\sigma$-model with torsion
by Hull and Spence (1989, 1991), Jack \etal\ (1990),
and R\v{o}cek and Verlinde (1992); the generalization
to $(p,q)$ supersymmetry was given by Hull, Papadopolous and Spence (1991);
the case of $N=2$ supersymmetry in superspace was discussed by 
R\v{o}cek and Verlinde (1992);
and the gauging of heterotic $\sigma$-models was discussed by
Hull (1994).  We will rederive these 
results, and include the effect of quantum corrections arising at
order $\OO(\apr^{0})$.  $J_{-}$ should be a gauge current and not
a complex structure leading to $(2,2)$ supersymmetry
(we will discuss the latter possibility in the next section), so we
will demand that $J_{-}$ be the top component of a supermultiplet,
and we will find its dimension $(0,1/2)$ superpartner $j_{-}$.

We begin by examining the variation of the action due to
the transformations given in equations \pref{ljphi}-\pref{ljlam}.
The bosonic terms in the variation
of the action are:
\begin{eqnarray}
\lefteqn{\delta S_{{\rm bos}} = \int d^{2}\sigma \left\{
	\p_{+}\epsilon \left[v^{\mu}\left(g_{\mu\nu}+b_{\mu\nu}\right)
	\p_{-}\phi^{\nu}
	\right]\right.}\nonumber \\
	&&\ \ \ + \p_{-}\epsilon \left[v^{\mu}
		\left(g_{\mu\nu}-b_{\mu\nu}\right)\p_{+}\phi^{\nu}
		\right] \nonumber \\
	&&\ \ \
	\left.+ \epsilon\left[ \p_{+}\phi^{\left(\mu\right. }\p_{-}\phi^{
		\left. \nu
		\right)}
	2\del_{\mu}v_{\nu} + 
	\p_{+}\phi^{\left[\mu\right.}\p_{-}\phi^{\left.\nu\right]} \left(
		\LL_{v}b_{\mu\nu}\right)
	\right]\right\} \ , \label{u1bosvar}
\end{eqnarray}
where $\del$ is the torsionless covariant derivative and $\LL_{v}$
is the Lie derivative with respect to $v$.  The vanishing of
the first term of the last line of \pref{u1bosvar} means that $v$
must be a Killing vector.
The second term in the last line of \pref{u1bosvar} should combine
with the first and second lines in a way that leaves only 
$\p_{+}J_{-}$.
If we let
\begin{equation}
\LL_{v}b = - d\omega\ ,
\label{lieb}
\end{equation}
then \pref{u1bosvar} becomes:
\begin{eqnarray}
\lefteqn{\delta S_{{\rm bos}} = \int d^{2}\sigma \left\{
	\p_{+}\epsilon \left[v^{\mu}\left(g_{\mu\nu}+b_{\mu\nu}\right)
	\p_{-}\phi^{\nu}
  - \frac{1}{2}\omega_{\mu}\p_{-}\phi^{\mu}\right]
		\right.}\nonumber \\
	&&\ \ \ \left. + \p_{-}\epsilon \left[v^{\mu}
		\left(g_{\mu\nu}-b_{\mu\nu}\right)\p_{+}\phi^{\nu}
 		+ \frac{1}{2} \omega_{\mu}\p_{+}\phi^{\mu}
		\right]\right\}\ . \label{u1bosvar2}
\end{eqnarray}
We wish the second line to vanish, so
\begin{equation}
\omega_{\mu} = -2v^{\rho} \left(g_{\rho\mu} - b_{\rho\mu}\right)\ .
\end{equation}
The first term in \pref{u1bosvar2} is then a piece of
$\p_{+}\epsilon J_{-}$, as expected by Noether's theorem.
Since $v$ is a Killing vector, Equation \pref{lieb} can be used to
show that:
\begin{equation}
\del^{(-)}_{\lambda}v^{\rho} = 0\ .
\label{delminusv}
\end{equation}
Combining this equation and Equation \pref{cscons}
we find that:
\begin{equation}
\LL_{v}J^{\mu}{}_{\nu} = 0\ .
\end{equation}
Equation \pref{lieb} also means that
\begin{equation}
\LL_{v}H = 0 \label{lieh}
\end{equation}
to order $\OO(\apr^{0})$.

The only term term in $\delta S$ quadratic in fermions is:
\begin{eqnarray}
\lefteqn{\delta S_{\lambda\lambda} = \int d^{2}\sigma \left\{ \right.
	\p_{+}\epsilon \Sigma_{ab}\lambda^{a}\lambda^{b} }\nonumber \\
	&&\left. + \epsilon\left[\frac{i}{2}v^{\rho}
	F_{\rho\lambda ab} + \hat{D}_{\lambda}\Sigma_{ab}
	\right]\p_{-}\phi^{\lambda}\lambda^{a}\lambda^{b}\right\} \ .
\end{eqnarray}
The first term is what we expect from Noether's theorem.
The vanishing of the second term
will be discussed below.
The final term in the variation of the action is quartic:
\begin{equation}
\delta S_{\psi\psi\lambda\lambda}
	= \int d^{2}\sigma
	\left[\frac{1}{2}v^{\rho}\hat{D}^{(+)}{}_{\rho}
	F_{\mu\nu ab} -2 i F_{\mu\nu c b}
		\Sigma^{c}{}_{a} \right]
	\psi^{\mu}\psi^{\nu}\lambda^{a}\lambda^{b}\ .
\end{equation}
Again, we will discuss the vanishing of this term below.

\figloc4{The one-loop $\OO(\apr^{0})$ contribution to
$\{J_{-},J_{-}\}$.}

Next we wish to impose the vanishing of the commutator of the $U(1)$
current with itself.  The classical bracket gives us:
\begin{equation}
\left\{ J_{-}(\sigma),J_{-}(\sigma')\right\}
	= -2v^{\mu}v_{\mu}\delta '(\sigma-\sigma')
	- 2 g_{\alpha\beta}v^{\alpha}\del_{\lambda}v^{\beta}
		\p_{-}\phi^{\lambda}\delta(\sigma-\sigma')\ .
\end{equation}
The antisymmetry of $H$ allows us to turn $\del$ into
$\del_{(-)}$ in the second term, which vanishes.  If
the $U(1)$ is realized entirely as a spacetime isometry then
this isometry must be null.  However, if we have a piece of
$\Sigma$ which scales as $1/\sqrt{\apr}$, then at order $\OO(\apr^{0})$
we get a one-loop correction to the above which comes from the
commutator of $\Sigma\lambda\lambda$ with itself.  We compute this
using the same strategy (and keeping in mind the same caveats) as the
previous section.
The relevant diagram is shown in Figure 4.  Converting
this term into a commutator expression gives us the
expression for the order $\OO(\apr^{0})$ part of the commutator:
\begin{eqnarray}
\lefteqn{
\left\{J_{-}(\sigma),J_{-}(\sigma')\right\}
= -2 \left( v^{\rho}v_{\rho} + \apr \Sigma^{ab}\Sigma_{ab} \right)
	\delta '(\sigma-\sigma')} \nonumber \\
	&&
	- \apr \Sigma^{ab}\hat{D}_{\lambda}\Sigma_{ab}\p_{-}\phi^{\lambda}
	\delta(\sigma-\sigma')\ .
\end{eqnarray}
to $\OO(\apr^{0})$ the final term vanishes, so that
\begin{equation}
v^{\rho}v_{\rho} + \apr \Sigma^{ab}\Sigma_{ab} = 0
\label{schwingvanish}
\end{equation}
if the $U(1)$ current is to be anomaly-free.

\figloc5{The one-loop $\OO(\apr^{0})$ contribution to
$\{G_{-},J_{-}\}$.}

Supersymmetry requires that the Poisson bracket of the 
supersymmetry current with
the top component $\phi_{1}$ of a superfield is:
\begin{equation}
\left\{ G_{-}(\sigma), \phi(\sigma')_{1} \right\}
	= - 2 h \phi_{0} \delta '(\sigma - \sigma')
	- \p_{-} \phi_{0} \delta (\sigma - \sigma')\
\end{equation}
where $h$ is the left conformal dimension of $\phi_{1}$
and $\phi_{0}$ is its superpartner, with left conformal dimension
$h-1/2$.
The classical Dirac bracket of the left-moving supersymmetry current with the
left-moving $U(1)$ current is:
\begin{eqnarray}
\lefteqn{ \left\{G_{-}(\sigma),J_{-}(\sigma') \right\}_{{\rm tree}}
	=  -2 \left(v^{\mu}e_{\mu a}\lambda^{a}\right)(\sigma')
		\delta' (\sigma - \sigma') + } \nonumber \\
	&&\ \ \ \left[ \left(2i \Sigma^{k}{}_{a} e_{\rho k}
		+ H_{\rho\gamma\mu}e^{\gamma}{}_{a}v^{\mu}\right)
		\p_{-}\phi^{\rho}\lambda^{a}
		- \p_{-}\left(v^{\mu}e_{\mu a}\lambda^{a}\right)\right.
		- \nonumber \\
	&&\ \ \	\left.\left(
		iv^{\mu}\hat{D}_{\mu}f_{abc}
		+ 6 f_{kab}\Sigma^{k}{}_{c}\right)\lambda^{a}\lambda^{b}
			\lambda^{c}\right]
		(\sigma')\delta(\sigma-\sigma')\ . \label{gcommJ}
\end{eqnarray}
In addition, the one loop part of the commutator of
$\Sigma^{(\perp)}\lambda\lambda$ (where $\Sigma^{(\perp)}$ is the piece of
$\Sigma$ scaling as $1/\sqrt{\apr}$)
and $f^{(\perp)}\lambda\lambda\lambda$
will contribute an order $\OO(\apr^{0})$ term.  This term will
come from the diagram shown in Figure 5.
Using the fact that
$\hat{D}\Sigma$ and $\hat{D}f^{(\perp)}$ are order $\OO(\apr^{0})$ or
smaller, to this order the one contribution at order $\OO(\apr^{0})$
is:
\begin{eqnarray}
\lefteqn{ \left\{ - i f^{(\perp)}{}_{abc}\lambda^{a}\lambda^{b}\lambda^{c},
\Sigma^{(\perp)}{}_{kl}\lambda^{k}\lambda^{l}\right\}_{{\rm 1 loop}}
	= - 6 i \apr f^{(\perp)}_{abc}\Sigma^{(\perp)bc}\lambda^{a}
	(\sigma')\delta '(\sigma-\sigma')} \nonumber \\
	&&\ \ \ - 3 i \apr f^{(\perp)}_{abc}\Sigma^{(\perp)bc}
	\left(\p_{-}\lambda^{a} + \hat{A}_{\mu}{}^{a}{}_{c}
		\p_{-}\phi^{\mu}\lambda^{c}\right)
		(\sigma')\delta(\sigma-\sigma')\ .
\end{eqnarray}
Using the fact that $\hat{D}f$ and $\hat{D}\Sigma$
vanish to this order, one can see that the last line is equal to:
\begin{equation}
	- 3 i \apr \p_{-}
		\left(f^{(\perp)}_{abc}\Sigma^{(\perp)bc}\lambda^{a}\right)\ ,
\end{equation}
as required by supersymmetry.  Thus, if
\begin{equation}
	e_{\rho k}\Sigma^{k}{}_{a}
		= \frac{i}{2}
			H_{\rho\alpha\mu}e^{\alpha}{}_{a}v^{\mu}\ ,
\end{equation}
and if the last line of Equation \pref{gcommJ} vanishes,
then $J_{-}$ is the top component of a superfield and its dimension
$(0,1/2)$ superpartner is:
\begin{equation}
j_{-} = \left( v^{\mu}e_{\mu a} + 3 i \apr f^{(\perp)}{}_{abc}
		\Sigma^{(\perp) bc}\right)\lambda^{a}\ .
\end{equation}

We also want to check that the anticommutator of $j$ with itself vanishes.
At order $\OO(\apr^{0})$ the tree level commutator will suffice.  The
condition for the vanishing of this anticommutator is:
\begin{equation}
	v^{\mu}v_{\mu} - 9 f_{akl}f^{amn}\Sigma^{kl}\Sigma_{mn} = 0\ .
\label{janom}
\end{equation}
If we write
\begin{equation}
\Sigma_{ab} = 3 i w^{c}f_{cab}\ ,\label{sigmatow}
\end{equation}
then Equation \pref{janom} follows from the nilpotency of $J_{-}$.

Let us summarize what we have found so far:
\begin{eqnarray}
&&\LL_{v}g_{\mu\nu} = 0 \label{lieg}\\
&&\LL_{v}H_{\mu\nu\rho} = 0 \\
&&\del_{(-)\lambda} v^{\rho} = 0 \label{delv} \\
&&\hat{D}_{\lambda}\Sigma_{ab} = 
	-\frac{i}{2}v^{\rho}F{}_{\rho\lambda ab} \label{dsig} \\
&&\frac{1}{2}v^{\rho}\hat{D}_{(+)\rho}
	F_{\mu\nu ab} = 2i\Sigma^{k}{}_{a}F_{\mu\nu kb} 
		\label{delcurve} \\
&& \left(1 - \PP \right)^{k}{}_{a}\Sigma_{kb} = 
	 \frac{i}{2}e^{\alpha}{}_{a}e^{\beta}{}_{b}v^{\rho}
		H_{\alpha\beta\rho} \label{tansig} \\
&&v^{\rho}\hat{D}_{\rho}f_{abc} = 6 i \Sigma_{k\left[ a\right.}
	f^{k}{}_{\left. bc\right]} \label{delf} \\
&&v^{\rho}{}v_{\rho} + \apr\Sigma^{ab}\Sigma_{ab} = 0 \ .
\end{eqnarray}
Not all of these equations are independent.  The tangent part of Equation
\pref{dsig} can be shown with some work to follow from Equations
\pref{tansig} and \pref{delv}.  The tangent part of
Equation \pref{delf} can be shown to be equivalent to Equation
\pref{lieh}. The tangent part of Equation \pref{delcurve}
is equivalent to:
\begin{equation}
\LL_{v} R_{(+)\mu\nu\alpha\beta} = 0\ .
\label{liecurve}
\end{equation}
The normal parts of Equations \pref{delcurve} and \pref{delf} take a
a little more thought.  $f$ and $\hat{A}$ have order $\OO(\sqrt{\apr})$
pieces, and $\Sigma$ may as well.  These will mix with order
$\OO(1/\sqrt{\apr})$ pieces of the currents to produce 
terms of $\OO(\apr^{0})$.

Let us break up $\Sigma$ into an order $1/\sqrt{\apr}$
piece $\Sigma^{(\perp)}$, an order $\apr^{0}$ piece
$\Sigma^{(0)}$, and an order $\sqrt{\apr}$ piece $\Sigma^{(1)}$.
$\Sigma^{(\perp)}$ and $\Sigma^{(1)}$ break up naturally into
normal and tangent pieces; Equation \pref{tansig} forces $\Sigma^{(\perp)}$
to be purely normal and fixes the tangent part of $\Sigma^{(0)}$.
Now, one may use the vanishing of either Equation
\pref{glrcomm} or the second line of Equation \pref{threelam}),
to rewrite Equation \pref{delf} as:
\begin{equation}
\lambda^{a}\lambda^{b}\lambda^{c}\left(
	\frac{1}{2}e^{\alpha}{}_{a} v^{\rho}F_{\alpha\rho b c}
	- 6 i \Sigma^{k}{}_{a} f_{k b c}\right) = 0\ .
\label{delf2}
\end{equation}
$(\Sigma^{(\perp)} + \Sigma^{(0)k}{}_{a})f^{(\perp)}{}_{kbc}$
has all normal indices and so cannot cancel any piece of the first term
in Equation \pref{delf2}; thus
\begin{equation}
\Sigma^{(\perp)}{}_{k\left[ a\right.}f^{(\perp)k}{}_{\left. bc\right]} 
= \Sigma^{(0)}{}_{k\left[ a\right.}f^{(\perp)k}{}_{\left. bc\right]} = 0\ ,
\label{sigmaftozero}
\end{equation}
where each equation follows from a different order of $\sqrt{\apr}$ in
Equation \pref{delf2}.  If $\PP^{a}{}_{k}\Sigma^{k}{}_{b}$
are written as in Equation \pref{sigmatow}, then Equation
\pref{sigmaftozero} follows from the Jacobi identity.
The $\OO(\apr^{0})$
piece $\Sigma^{(\perp)k}{}_{a}f^{(1)}{}_{kbc}$ has two tangent indices
and one normal index and cannot cancel off any piece of the first term of
Equation \pref{delf2}.  Using Equation \pref{fcorr}, we find that
\begin{equation}
F_{\alpha\beta kl}f^{(\perp)akl}\Sigma^{(\perp)}{}_{ab} = 0\ .
\label{curvesigvan}
\end{equation}
If we let $\Sigma^{(\perp)}_{ab} = f^{(\perp)}_{abc}w^{c}$ and
$F_{\mu\nu ab} = F^{(c)}_{\mu\nu}f^{(\perp)}{}_{abc}$, then this
equation implies that
\begin{equation}
F_{\mu\nu}{}^{a}{}_{b}w^{b} = \left[\hat{D}_{\mu},\hat{D}_{\nu}
	\right]^{a}{}_{b}w^{b} = 0 \ .
\end{equation}
The last line follows at this order from the fact that
$\hat{D}\Sigma^{(\perp)}$ and $\hat{D}f$ also vanishes to lowest order;
corrections will come at order $\OO(\sqrt{\apr})$.  The final
$\OO(\apr^{0})$ piece of Equation \pref{delf2} comes from
$\Sigma^{(1)k}{}_{a}f^{(\perp)}{}_{kbc}$; thus Equation \pref{delf2}
requires that:
\begin{equation}
\Sigma^{(1)c}{}_{a} = \frac{3i}{4} \apr 
	e^{\alpha}{}_{a}v^{\beta}F{}_{\alpha\beta kl}
	f^{(\perp)ckl}\ .
\label{sigmaimprove}
\end{equation}
Note that higher order corrections to $J_{-}$ and to $G_{-}$ will
also lead to higher order corrections to $j_{-}$.

The right hand side of Equation \pref{delcurve} has an
$\OO(1/\sqrt{\apr})$ piece $\Sigma^{(\perp)k}{}_{a}F_{\mu\nu kb}$,
and two order $\apr^{0}$ pieces, one from $\Sigma^{(0)}F$
and one from $\Sigma^{(\perp)}F^{(1)}$,
where $F^{(1)}$ is the order $\OO(\sqrt{\apr})$ piece of $F$
arising from Equation \pref{conncorr}.  This last piece has one normal
and one tangent gauge index, while the first piece and the left-hand side
of Equation \pref{delcurve} have two normal indices and so vanish
separately.  The vanishing of the term with mixed indices, 
\begin{equation}
\Sigma^{k}{}_{a} \left( \hat{D}_{\mu}\hat{A}^{(1)}{}_{\nu}{}^{l}{}_{b}
	G_{kl} - \hat{D}_{\nu}\hat{A}^{(1)}{}_{\mu}{}^{l}{}_{b}G_{kl}
	\right)\ ,
\end{equation}
follows from Equations \pref{conncorr} and \pref{curvesigvan}
since the covariant derivatives of $\Sigma$, $\PP$, and
$f^{(\perp)}$ will show up at order $\sqrt{\apr}$.
Using Equation \pref{dsig} and the fact that
\begin{equation}
\left[\hat{D}_{\mu},\hat{D}_{\nu}\right]^{a}{}_{k}\Sigma^{k}{}_{b}
	= F_{\mu\nu}{}^{a}{}_{k}\Sigma^{k}{}_{b}
	- F{}_{\mu\nu}{}^{l}{}_{b}\Sigma^{a}{}_{l}\ ,
\end{equation}
the rest of Equation \pref{delcurve} can be reduced to the 
$\OO(\apr^{0})$ part of the Bianchi identity:
\begin{equation}
	\hat{D}^{(0)}_{\left[\rho\right.}F^{(\perp)(0)}_
	{\left.\mu\nu\right]}{}^{a}{}_{b}
	= 0\ .
\end{equation}		

In summary, the {\em independent} constraints arising from the $U(1)$
symmetry are that the worldvolume geometry admit an isometry
(Equations \pref{lieg} and \pref{lieh}) which is
covariantly conserved (Equation \pref{delv}).  The
part of the $U(1)$ current which is quadratic in $\lambda$ must satisfy
Equations \pref{dsig} and \pref{tansig}.  The isometry and this
quadratic term are related by the requirement that the
anomaly vanish, Equation \pref{schwingvanish}

Note that we could have started with the most general dimension
$(0,1/2)$ operator $j_{-}$ and then found its dimension $(0,1)$
superpartner $J_{-}$.  If we define $v^{a}$ so that
\begin{eqnarray}
	&&(1-\PP)^{a}{}_{b}v^{b} = e^{a}{}_{\mu}v^{\mu} \\
	&&\PP^{a}{}_{b}v^{b} = w^{b}
\end{eqnarray}
then $j_{-}$ will have the simple form
\begin{equation}
j_{-} = v_{a}\lambda^{a}\ .
\end{equation}
However, by starting with the most general dimension $(0,1)$ current
many of the results of this section can be used to find
constraints on the geometry necessary for global $(2,2)$ and $(2,4)$
supersymmetry.  (Recall that since the right-moving sector has
$\hat{c}=4$ and $N=2$ supersymmetry, it automatically
has global $(4,1)$ supersymmetry as well: see Eguchi \etal\ (1989)).
We will not derive these constraints here, but let us outline the
necessary calculations.  For $(2,2)$ supersymmetry
one would start by demanding that the first line of Equation
\pref{gcommJ} vanish so that $J_{-}$ was the bottom component of
a superfield; the top component would be the additional supersymmetry charge
$G^{(2)}_{-}$.
Note that in this calculation $\Sigma$ would scale as $\apr^{0}$.  The
conditions for invariance of the action have already been worked out;
in order to find the rest of the constraints one would need to
ensure closure
of the $N=2$ supersymmetry algebra (including the condition that
$J_{-}$ defines a $U(1)$ current algebra with level $c/3$).
For $(2,4)$ supersymmetry one would need to find two other 
dimension $(0,1)$
operators in order to make up an $SU(2)$ current algebra; their
dimension $(0,3/2)$
superpartners would be the remaining supersymmetry currents
of the $N=4$ algebra.

\section{Physical and geometric interpretation}
	
The interpretation of the $N=(2,1)$ theory as a mapping of
a $2+2$-dimensional worldvolume into some spacetime is not obvious.
In particular, we should not directly identify
the vector bundle $\VV$ with the spacetime or its tangent space.
We can see this by thinking about examples with target space
supersymmetry.  In these cases the fermions living in the normal part
of $\VV$ are grouped into 8 groups of $SU(2)$ triplets
and so $\VV$ is broken up into 3 dimensional subspaces.
In each of these subspaces two of the fermions are bosonized;
the boson couples to a background $U(1)$ field which we will argue
is related to a coordinate of the worldvolume in spacetime,
while the fermion is its partner under worldsheet supersymmetry.
We should not identify the chiral boson directly with a
spacetime coordinate, just as one does not give the internal
sector of the usual $(1,0)$ heterotic string a spacetime interpretation;
the chiral boson lives on a circle with a fixed radius and there
is no graviton operator in the physical spectrum that
would change this radius.

Instead, the spacetime should be directly
related to the configuration space of the target space fields,
in analogy to the soliton string constructions of
Harvey and Strominger (1995) and Sen (1995);
in these constructions the soliton string is the fundamental string
of the dual theory, and the dual spacetime
is the moduli space of zero-mode fluctuations
around the soliton solution.  In the $N=(2,1)$ constructions
the translation of this statement is that
the spacetime of the target space string or membrane is
parameterized for small fluctuations
by the expectation value of the massless vertex operators
of the $(2,1)$ string (Kutasov and Martinec 1996;
Kutasov, Martinec and O'Loughlin 1996).

In these $(2,1)$ constructions, the
vertex operators corresponding to string or membrane
excitations are those for the Yang-Mills fields and
for worldvolume gravity.  We can see how the
gauge field excitations are realized as coordinate
excitations by writing
\begin{equation}
\hat{A}_{\mu} = J^{\lambda}{}_{\mu} h^{-1}\p_{\lambda}h \ .
\label{atoscalar}
\end{equation}
Here
\begin{equation}
h = e^{\phi^{a} t^{a}}\ ,
\end{equation}
where $\phi^{a}$ are scalars living in the adjoint of the
gauge group and $t$ are Hermitian generators of the Lie
algebra, in the adjoint representation.  Equation \pref{atoscalar}
solves the constraint \pref{comcurve} (one may show this
using Equation \pref{nienhuis}); in complex coordinates
\pref{atoscalar} means that $F_{ij}$ and $F_{\bar{i}\bar{j}}$
vanish.  The vertex operators constructed from the
currents of the internal gauge group
in fact represent fluctuations of $\phi^{a}$ (Ooguri and Vafa 1991b).
It seems natural to identify $\phi^{a}$ with the transverse
coordinates of the 4-volume; this would mean that the spacetime
is a group manifold.  In the case $G=U(1)^{8}$, the group required for
target space supersymmetry,
the gauge fields will be parameterized by 8 scalars
$\phi^{a}$; the gauge field strength will be
\begin{equation}
F^{a}{}_{\mu\nu} = 2 \p_{\left[\mu\right. }
	J^{\alpha}{}_{\left.\nu\right]}\p_{\alpha}\phi^{a}
	- 2 J^{\alpha}{}_{\left[ \mu\right. }
	\p_{\left.\nu\right]}\p_{\alpha}\phi^{a}\ .
\end{equation}
In complex coordinates, this reads:
\begin{equation}
F^{a}{}_{i \bar{j}} = -2 i \p_{i}\p_{\bar{j}}\phi^{a}\ .
\end{equation}

The left-moving supersymmetry imposes a nontrivial structure on $\VV$;
in the action \pref{cuferm} there is no relation between geometric
structures in $\VV$ and the intrinsic geometry of the target space.
This relation is encoded in the left-moving supersymmetry
current \pref{origlsusy}, and in particular in the form
$e_{\mu}{}^{a}$ which maps
the tangent bundle of the target space into $\VV$.
Equations \pref{cove} and \pref{reqf2} show that the geometric
structures of the target space and of the tangent part of $\VV$
are related as well.
Note that Equation \pref{cove} can be rewritten as:
\begin{equation}
\PP^{a}{}_{k}\hat{D}_{\rho}\PP^{k}{}_{b}
	= \frac{3}{2}\apr e^{\mu}{}_{b}F_{\rho\mu kl}
		f^{(\perp)akl}\ . \label{covp2}
\end{equation}
Although the details are not clear,
these equations should be related to equations describing imbedded
surfaces, such as the Gauss and Codazzi equations.\footnote{This 
interpretation was suggested by E. Martinec.}
The first equation relates the intrinsic worldvolume curvature to
the tangent part of $F$, and is reminiscent of the
Gauss equation
(the ``Theorem Egregium'').  The second equation relates $\hat{D}\PP$
to $F^{(\perp)}$; $F^{(\perp)}$ is described
by second derivatives of $\phi$ and could be related to the
second fundamental form of the surface.  There are many
issues which need to be resolved before we can construct a
definitive interpretation of the $(2,1)$ target
space theory.  One problem is that the normal and tangent gauge
fields do not seem to have the same status. 
The transverse gauge fields have a natural
interpretation as coordinates; the tangent gauge fields, however,
seem to be mapped into the spin connection of the 4-volume.
For example, if the gauge fields of the normal part of $\VV$
are valued in the Lie algebra of $U(1)^{8}$, the tangent
gauge fields have some non-Abelian structure induced from
the local Lorentz group acting on tangent frames of the target space;
it seems that the tangent gauge fields should not be related to
coordinates in the same way as the normal gauge fields.
Perhaps the $(2,1)$ $\sigma$ model
gives some sort of static gauge description of imbedded 4-volumes.

Supersymmetry seems to give us geometric constraints on the
imbedding of 4-volumes into spacetime;
the classical equations of motion for the target worldvolume
are simply be the $\beta$-function equations for the heterotic string
(Callan \etal\ 1985; Fradkin and Tseytlin 1985b; Lovelace 1986;
Sen 1986; Hull 1986a; Bonneau and Valent 1994).  These equations seem
to describe some sort of coupling of the self-dual Yang-Mills equations
to self-dual gravity (Ooguri and Vafa 1991a,b).
To see how these equations might be related to the equations
of motion for the 4-volume, let us turn off gravity.
The self-duality condition on $F$ is:
\begin{equation}
\epsilon^{\mu\nu\lambda\rho}J_{\mu\nu}
	F_{\lambda\rho}{}^{a}{}_{b} = 0\ .
\end{equation}
If we use the ans\"{a}tz \pref{atoscalar} and work in
complex coordinates, then these equations become
(Nair and Schiff 1990,1992; Ooguri and Vafa 1991b):
\begin{equation}
g^{i\bar{j}}\p_{i}\left(h^{-2}\p_{\bar{j}}h^{2}\right)=0\ .
\end{equation}
In the Abelian theory this equation reduces to:
\begin{equation}
\left(\p_{1}\p_{\bar{1}}-\p_{2}\p_{\bar{2}}\right)\phi^{a}=0\ ,
\end{equation}
which is just the equation of motion for a free scalar field
in $R^{(2,2)}$.

It is worth noting that one may phrase the classical equations
of motion and constraints derived from the Nambu action for $p$-branes
in terms of the Gauss and Codazzi equations and the vanishing of the
trace of the second fundamental form of the imbedded
worldvolume (Bandos \etal\ 1995).

\section{Conclusion}

We have clearly taken only a small, if necessary, step towards understanding
the classical and quantum dynamics of the target space of $N=(2,1)$
strings.  However, we can begin to see what questions to ask.

At present there is no obvious way to describe the imbedding of
these world volumes in a more general, curved spacetime.  One
possibility comes from the fact that the self-dual Yang-Mills
equations are equivalent to the classical equations of motion
for a four dimensional generalization of the WZW model
(Nair and Schiff 1990, 1992).  This model
looks like a 4-dimensional $\sigma$-model with an extra Wess-Zumino
term, where the $\sigma$-model fields live in a group manifold.  
There is some belief that this theory exists as a quantum
theory and has some current algebra structure
similar to the 2-dimensional WZW model (Losev \etal\ 1995).
If we could understand better if and why this model exists,
and how we might couple it to the gravity sector of the
$N=(2,1)$ target space,
we might be able to generalize the WZW$_{4}$
model to more general target
spaces.

We might also hope to make contact with other recent suggestions
of $2+2$-dimensional worldvolumes in $10+2$-dimensional spacetimes.
Many results have been
obtained by compactifying F-theory on restricted classes
of K3 surfaces (Vafa 1996), Calabi-Yau threefolds (Morrison and Vafa 1996),
and other manifolds (Witten 1996a,b); one can make statements about
the theory in curved space, unlike the theory presented in this paper.
On the other hand,
it is not clear what the ``fundamental'' objects are; there are only
some suggestions, discussed in the Introduction, that 
they might be four dimensional.
Furthermore, the principles for constructing F-theory vacua are
unknown; the aforementioned results for compactifications
of F-theory come from comparing the moduli space of the 
compactification manifold with the moduli space of other theories.
It would be nice if there was a direct connection between F-theory and the
$(2,1)$ string, as there are some analogous structures, but how to
make this link is not obvious.

It may be that if and when we successfully define a perturbation theory
of sums over 4-volumes, that this theory 
will break down at high energies and large
orders just as string theory seems to.  This is no reason not to go forward;
string theory has done remarkably well in providing hints of 
its underlying structure, and has even given us enough information
(by constraining the low-energy vacua) to find
some of its nonperturbative physics.  Perhaps this theory of 4-volumes
will do the same.

\vskip .4cm
\leftline{\bf \large Acknowledgments}
\vskip .3cm

I would like to thank E. Martinec for
suggesting this problem, and for many invaluable discussions,
explanations, suggestions, and insights.  
In particular, Section 6
arose largely from 
discussions with him.  M. O'Loughlin also deserves special thanks for
innumerable useful conversations.
I also would like to acknowledge helpful
conversations with J. Harvey, E. Poppitz,
and L. Yaffe.  This paper is being submitted in partial fulfillment
of the requirements for the PhD degree at the University of Chicago.

\appendix

\section{Target space and worldsheet geometry}

\subsection{Target space geometry}

In general the target space of the $N=(2,1)$ string will have
a metric $g_{\mu\nu}$, with signature $(2,2)$, and an antisymmetric
2-form $b_{\mu\nu}$ from which we may form the 3-form field strength
\begin{equation}
	H_{\mu\nu\lambda} = \p_{\mu}b_{\nu\lambda} 
	+ \p_{\nu} b_{\lambda\mu} + \p_{\lambda} b_{\mu\nu}\ ,
\end{equation}
or equivalently,
\begin{equation}
	H = 3 d b\ .
\end{equation}
$H$ is a torsion tensor:
we define the torsionful connection as:
\begin{equation}
{\Gamma_{(\pm)}}^{\mu}_{\nu\lambda} =
	{\Gamma_{(c)}}^{\mu}_{\nu\lambda} \pm \frac{1}{2} H^{\mu}_{\nu\lambda}
\end{equation}
where $\Gamma_{(c)}$ is the Christoffel connection.
Covariant derivatives with torsion are defined as usual:
\begin{eqnarray}
\del_{(\pm)\lambda} K^{\mu\nu\ldots}{}_{\alpha\beta\ldots}
	&=& \p_{\lambda} K^{\mu\nu\ldots}{}_{\alpha\beta\ldots}
	+ {\Gamma_{(\pm)}}^{\mu}_{\lambda\rho} 
	K^{\rho\nu\ldots}{}_{\alpha\beta\ldots}
	+ {\Gamma_{(\pm)}}^{\nu}_{\lambda\rho} 
	K^{\mu\rho\ldots}{}_{\alpha\beta\ldots} + \ldots \nonumber \\
	&& - {\Gamma_{(\pm)}}^{\rho}_{\lambda\alpha}
	K^{\mu\nu\ldots}{}_{\rho\beta\ldots}
	- {\Gamma_{(\pm)}}^{\rho}_{\lambda\beta}
	K^{\mu\nu\ldots}{}_{\alpha\rho\ldots} + \ldots\ .
\end{eqnarray}
The curvature with torsion is defined via the equation
\begin{equation}
\left[\del_{(\pm)\mu},\del_{(\pm)\nu}\right] v^{\lambda} = 
	{R_{(\pm)}}^{\lambda}{}_{\rho\mu\nu} v^{\rho}
	\mp H^{\alpha}{}_{\mu\nu} \del_{(\pm)\alpha}v^{\lambda} ,
\end{equation}
and it can be written in terms of the connection as:
\begin{equation}
	{R_{(\pm)}}^{\lambda}{}_{\rho\mu\nu} = 
	\p_{\mu} {\Gamma_{(\pm)}}^{\rho}{}_{\nu\lambda}
	+ {\Gamma_{(\pm)}}^{\rho}{}_{\mu\gamma}
	{\Gamma_{(\pm)}}^{\rho}{}_{\mu\gamma} -
	(\mu \longleftrightarrow \nu)\ .
\end{equation}
$R_{(\pm)}$ satisfies the following identities:
\begin{eqnarray}
	R_{(\pm)\mu\nu\alpha\beta} &=& - R_{(\pm)\nu\mu\alpha\beta}
	= - R_{(\pm)\nu\mu\alpha\beta} = - R_{(\pm)\nu\mu\beta\alpha} \\
	R_{(+)\mu\nu\alpha\beta} &=& R_{(-)\alpha\beta\mu\nu}\ .
\end{eqnarray}

The left-moving fermions live on a 28-dimensional vector bundle $\VV$
with signature $(26,2)$ and metric $G_{ab}$.  Generally, in each fiber
we split the space into a 4-dimensional space with signature $(2,2)$
and an orthogonal Euclidean 24-dimensional space.
This gives us two separate vector bundles which
are complements in $\VV$.
The bundle of 4-dimensional vector spaces will be
identified with the tangent bundle of the target space,
while the other 24 will live in some internal
vector bundle fibered over the target spacetime.  If ${A_{\mu}}^{a}{}_{b}$ is
an antisymmetric connection on this bundle, then the
metric-compatible connection is:
\begin{equation}
{\hat{A}_{\mu}}{}^{a}{}_{b} = {A_{\mu}}^{a}_{b} + \frac{1}{2}
	G^{ac}\p_{\mu}G_{cb}\ ,
\end{equation}
which is not antisymmetric.
In this paper we use the covariant derivative:
\begin{eqnarray}
\lefteqn{\hat{D}_{\lambda} M^{ab\ldots}{}_{cd\ldots} = 
	\p_{\lambda} M^{ab\ldots}{}_{cd\ldots}
	+ \hat{A}_{\mu}{}^{a}{}_{k} M^{kb\ldots}{}_{cd\ldots} +}
		\nonumber \\
	&&\hat{A}_{\mu}{}^{b}{}_{k} M^{ak\ldots}{}_{cd\ldots} + \ldots -
	\hat{A}_{\mu}{}^{k}{}_{c} M^{ab\ldots}{}_{kd\ldots}-
	\hat{A}_{\mu}{}^{k}{}_{d} M^{ab\ldots}{}_{ck\ldots}- \ldots
\label{gcov}
\end{eqnarray}
$\hat{D_{\pm}}$ is covariantized with respect to both tangent space
and vector bundle indices in the obvious way.
The field strength that appears in the action is
\begin{equation}
{F_{\mu\nu}}{}^{a}{}_{b} = \p_{\mu} {\hat{A}_{\nu}}{}^{a}{}_{b}
	+ {\hat{A}_{\mu}}{}^{a}{}_{c} {\hat{A}_{\nu}}{}^{c}{}_{b}
	- (\mu \longleftrightarrow \nu)\ .
\end{equation}
We raise and lower indices directly on this field strength tensor
rather than defining $F_{\mu\nu ab}$ as 
$\p_{\mu}\hat{A}_{\nu ab}(+\ldots)$; with some work one can show that
$F_{ab}$ is antisymmetric in the gauge indices.
Note that unlike the field strength defined by Hull and Witten (1985),
here $F$
is in fact the commutator of gauge covariant derivatives:
\begin{equation}
\left[ \hat{D}_{\mu},\hat{D}_{\nu} \right]v^{a} =
	F_{\mu\nu}{}^{a}{}_{b}v^{b}\ .
\end{equation}
We may also work in the tangent space of $\VV$
using the
vielbein $\rho^{A}_{a}$, where
\begin{equation}
	\eta_{AB}\rho^{A}_{a}\rho^{B}_{b} = G_{ab}\ , \ 
	G^{ab} \rho^{A}_{a}\rho^{B}_{b} = \eta^{AB}
\end{equation}
Here lowercase indices $(a,b,...)$ are vector bundle indices,
raised and lowered with the
curved metric $G$ and uppercase indices are the indices for
the tangent space to the bundle with the flat metric $\eta$.
Tensors with uppercase indices can be converted to tensors with
lowercase indices by contraction with $\rho$.
The covariant derivatives are as in Equation \pref{gcov},
with the lowercase indices replaced by uppercase indices and
the connection $\hat{A}$ replaced by the antisymmetric
connection one-form
\begin{equation}
	{\omega_{\mu}}_{AB} = \hat{A}_{\mu a b} \rho^{a}_{A} \rho^{b}_{B}
	+ \rho_{A c} \p_{\mu} \rho^{c}_{B}\ .
\end{equation}

If we have a complex structure with a vanishing Nienhuis tensor
we may choose a coordinate system such that:
\begin{eqnarray}
&&J^{i}{}_{\bar{j}} = J^{\bar{i}}{}_{j} = 0 \\
&&J^{i}{}_{j} = i \delta^{i}{}_{j} \\
&&J^{\bar{i}}{}_{\bar{j}} = - i \delta^{\bar{i}}_{\bar{j}}\ .
\end{eqnarray}
In these coordinates the flat space metric in $2+2$ dimensions is:
\begin{equation}
ds^{2} = -dz^{1}dz^{\bar{1}} + dz^{2}dz^{\bar{2}}\ .
\end{equation}

\subsection{Worldsheet (super-) geometry}

Throughout this paper, we work in conformal gauge.  The worldsheet metric
is:
\begin{equation}
ds^{2} = e^{2\phi} \left(-d\tau^{2} + d\sigma^{2}\right)\ .
\end{equation}
We will be working on the cylinder, so
\begin{equation}
-\infty <\tau < \infty , \ 0 \leq \sigma \leq 2\pi\ .
\end{equation}
Light cone derivatives are defined
as:
\begin{equation}
\p_{\pm} = \frac{\p}{\p\tau} \pm \frac{\p}{\p\sigma}\ ,
\end{equation}
so that
\begin{equation}
\sigma_{\pm} = \frac{\tau\pm\sigma}{2}\ .
\end{equation}

The $(1,0)$ superfields are:
\begin{eqnarray}
\Phi^{\mu} &=& \phi^{\mu} + \theta_{+} \psi^{\mu} \\
\Lambda^{a} &=& \lambda^{a} + \theta_{+} F^{a}
\end{eqnarray}
where $F$ is an auxiliary field.  The superspace derivative is:
\begin{equation}
D_{+} = \frac{\p}{\p\theta_{+}} + i \theta_{+}\p_{+}\ .
\end{equation}
For $(2,0)$ superspace we follow the conventions of
Dine and Seiberg (1986).  There are two commuting Grassman coordinates 
$\theta_{+} , \theta^{\ast}_{+}$ paired with
$\sigma_{+}$.  The superspace derivatives are:
\begin{eqnarray}
D_{+} &=& \frac{\p}{\p\theta_{+}} + i \theta^{\ast}_{+}\p_{+} \\
D^{\ast}_{+} &=& \frac{\p}{\p\theta^{\ast}_{+}} + i \theta_{+}\p_{+} \\
\end{eqnarray}
Chiral and antichiral superfields are written in complex coordinates
with conjugate indices:
\begin{eqnarray}
D^{*}_{+}\Phi^{i} &=& 0 \,\Longrightarrow\,
	\Phi^{i} = \phi^{i} + \sqrt{2}\theta_{+}\psi^{i}
	- i \theta^{*}_{+}\theta_{+}\p_{+}\phi \\
D_{+}\Phi^{\bar{j}} &=& 0 \,\Longrightarrow\,
	\Phi^{\bar{j}} = \phi^{\bar{j}} + \sqrt{2} \theta^{\ast}_{+}
		\psi^{\bar{j}} + i \theta^{\ast}_{+}\theta_{+}\p_{+}
			\phi^{\bar{j}}\ .
\end{eqnarray}

\section{Fundamental Poisson and Dirac brackets of the $N=(2,1)$
$\sigma$-model}

First let us recall the construction of Poisson brackets for 
systems with fermions (for a clear explanation see ch. 6,
sections 4 and 5 of Henneaux and Teitelboim (1992), from
which the discussion in this paragraph was lifted; we repeat this discussion
in order to explain our conventions).  Let us look at a system
with commuting phase space
coordinates and momenta $(q,p)$ and anticommuting coordinates and
momenta $(\theta, \pi)$; the latter define an obvious $\ZZ_{2}$ 
grading, so we can
define monomials and by extension certain functions
as being even or odd with respect to this grading.  
We define partial derivatives as acting from the
right; differentials are given by the formula
\begin{equation}
\delta F(z) = \delta z_{i} \frac{\partial F}{\partial z_{i}}\ .
\end{equation}
Given a Lagrangian $L(q,\dot{q},\theta,\dot{\theta})$,
the Hamiltonian as:
\begin{equation}
H = \dot{q} p + \dot{\theta}\pi - L
\end{equation}
If we set to zero the
infinitesimal variation of $H$ with respect to time translations,
we of course find Hamilton's equations:
\begin{eqnarray}
\dot{p} &=& - \frac{\partial H}{\partial q} \\
\dot{q} &=& \frac{\partial H}{\partial p} \\
\dot{\pi} &=& - \frac{\partial H}{\partial \theta} \\
\dot{\theta} &=& - \frac{\partial H}{\partial \pi}\ . \label{thdot}
\end{eqnarray}
Note the minus sign in Equation \pref{thdot}.  Now if $H$ is an even function
then we can compute
\begin{equation}
\frac{d f}{d t} = \frac{\partial f}{\partial t} + \left\{f, H\right\}
\end{equation}
where if $B$ is an even function 
the Poisson bracket above is defined as
\begin{equation}
\left\{A,B\right\} = \frac{\partial A}{\partial q^{i}}
\frac{\partial B}{\partial p_{i}} + 
\frac{\partial A}{\partial p_{i}}\frac{\partial B}{\partial q^{i}}
+ \frac{\partial B}{\partial \pi_{a}}\frac{\partial A}{\partial\theta^{a}}
+ \frac{\partial B}{\partial \theta^{a}}\frac{\partial A}{\partial\pi_{a}}
\label{pb1}\ .
\end{equation}
If $B$ is odd and $A$ is even then we define formally
\begin{equation}
\left\{A,B\right\} \equiv - \left\{B,A\right\}\ ,
\end{equation}
where the bracket on the right hand side of this equation is defined above.  
For $A$ and $B$ both odd,
we should be a little careful: if we want the bracket to be symmetric
and to have the associative properties of an anticommutator
\begin{equation}
\left\{AB, C\right\} = A\left\{B,C\right\} - \left\{A,C\right\} B\ ,
\end{equation}
then the signs in front of the last two terms of Equation \pref{pb1}
are reversed.  The final expression for the commutator of functions
$A$ and $B$ with definite grading
is (Henneaux and Teitelboim 1992):
\begin{equation}
\left\{A,B\right\} = \left(\frac{\partial A}{\partial q^{i}}
\frac{\partial B}{\partial p_{i}} + 
\frac{\partial A}{\partial p_{i}}\frac{\partial B}{\partial q^{i}}\right)
+ (-1)^{\sigma(A)}\left(
\frac{\partial A}{\partial \theta^{i}}\frac{\partial B}{\partial \pi_{i}}
+ \frac{\partial A}{\partial\pi^{i}}
	\frac{\partial B}{\partial\theta_{i}}\right)\ .
\end{equation}
where $\sigma(A)$ is $+1$ if $A$ is even and $-1$ if $A$ is odd
The canonical Poisson brackets are:
\begin{equation}
\{q,p\} = 1\ ; \ \{\theta,\pi\} = -1\ .
\end{equation}

If $\tau$ is chosen as the worldsheet time variable,
the canonical
momenta of the action \pref{ract}$+$\pref{cuferm} are:
\begin{eqnarray}
p_{\phi,\mu} &=& g_{\mu\nu}\dot{\phi}^{\nu} - b_{\mu\nu}\phi'{}^{\nu}
+ \frac{i}{2}g_{\alpha\rho}\Gamma_{(+)}{}^{\rho}{}_{\mu\beta}
	\psi^{\alpha}\psi^{\beta}
+ \frac{i}{2} A_{\mu ab} \lambda^{a}\lambda^{b} \\
\pi_{\psi,\mu} &=& - \frac{i}{2} g_{\mu\nu}\psi^{\nu} \\
\pi_{\lambda,a} &=& = - \frac{i}{2} G_{ab}\lambda^{b}\ .
\end{eqnarray}
This system is clearly constrained.  We follow Dirac's procedure
(Dirac 1967; see also Hanson, Regge and Teitelboim 1976 for
an introduction and many examples).  The constraints
\begin{eqnarray}
\chi_{\mu} &=& \pi_{\mu} + \frac{i}{2} g_{\mu\nu}\psi^{\nu} = 0\\
\chi_{a} &=& \pi_{a} + \frac{i}{2} G_{ab}\lambda^{a} = 0
\end{eqnarray}
are second class:
\begin{eqnarray}
\{\chi_{\mu},\chi_{\nu}\} &\equiv& C_{\mu\nu} = -i g_{\mu\nu} \\
\{\chi_{a},\chi_{b}\} &\equiv& C_{a b} = -i G_{a b}\ .
\end{eqnarray}
We may compute the Dirac brackets in standard fashion to find:
\begin{eqnarray}
&&\left\{ p_{mu}(\sigma), \phi^{\nu}(\sigma') \right\}_{D} = 
	\delta_{\mu}{}^{\nu}
	\delta (\sigma - \sigma') \\
&&\left\{ \psi^{\mu}(\sigma), \psi^{\nu}(\sigma') \right\}_{D} =
	- i g^{\mu\nu} \delta (\sigma - \sigma')\\
&&\left\{ p_{\mu},\psi^{\nu} \right\}_{D} = \frac{1}{2}g^{\nu\beta}
	g_{\beta\rho,\mu}\psi^{\rho}\delta (\sigma - \sigma')\\
&&\left\{ p_{\mu},\lambda^{a} \right\}_{D} = \frac{1}{2}G^{ad}
	G_{db,\mu}\lambda^{b} \\
&&\left\{ p_{\mu},p_{\nu} \right\}_{D} =
	- \frac{i}{4} g^{\alpha\beta}g_{\alpha\lambda,\mu}
	g_{\beta\nu,\rho} \psi^{\lambda}\psi^{\rho} +
	- \frac{i}{4} G^{ab}G_{ac,\mu}g_{bd,\nu}
	\lambda^{c}\lambda^{d}\ .
\end{eqnarray}
The last three brackets are due to the metric factor in the kinetic
term for $\psi$ and $\lambda$. If we rotate the fermions with
the appropriate vielbeins, so that the kinetic terms are the
standard flat-space terms, the last three brackets above
will vanish.  

In calculations
in the body of the paper, all brackets are Dirac brackets (so
we drop the subscript), and we lift
them to quantum commutators in the standard way.

\section{Conventions for worldsheet Green's functions}

In this paper we are interested in the singular short-distance
behavior of commutators and operator products.  For these purposes we
may assume that $\sigma$ extends over the real line.  
If the action in flat space
is normalized like so:
\begin{equation}
S = \frac{1}{4\pi\apr}\int d\sigma d\tau
	\left( \eta_{\mu\nu}\p_{+}\phi^{\mu}\p_{-}\phi^{\nu}
	+ i \eta_{ab}\lambda^{a}\p_{+}\lambda^{b} + i \eta_{\mu\nu}
	\psi^{\mu}\p_{-}\psi^{\nu}\right)\ ,
\end{equation}
then the $\phi$ propagator will be:
\begin{eqnarray}
\langle \phi^{\mu}(\sigma)\phi^{\nu}(\sigma') \rangle
	&=& 2\pi i\apr \eta^{\mu\nu}\int \frac{d^{2}k}{(2\pi)^{2}}
	\frac{e^{i k\cdot (\sigma-\sigma')}}{k^{2}+i\epsilon} \\
	&=& -\frac{\apr}{2} \eta^{\mu\nu}
		\ln(\sigma_{-}-\sigma'_{-})(\sigma_{+}-\sigma'_{+})\ ,
\end{eqnarray}
while the $\lambda$ propagator will be:
\begin{eqnarray}
\langle \lambda^{a}(\sigma_{-})\lambda^{b}(\sigma'_{-}) \rangle
	&=& 2\pi i\apr\eta^{ab}\int \frac{d^{2}k}{(2\pi)^{2}}
	\frac{k_{+}e^{i k\cdot (\sigma-\sigma')}}{k^{2}+i\epsilon} \\
	&=& \frac{-i\apr\eta^{ab}}{2(\sigma_{-}-\sigma'_{-})}\ .
\end{eqnarray}
and similarly for $\psi$, with $\sigma_{-},k_{+}\rightarrow \sigma_{+},k_{-}$.
If we wish to convert operator product singularities to equal-time
commutator singularities, we may use the formula:
\begin{equation}
\frac{1}{2(\sigma_{-}-\sigma'_{-})}\longrightarrow
	\delta (\sigma-\sigma')\ .
\end{equation}

\pagebreak
\leftline{\bf \Large \noindent References}
\vskip .5cm
\begin{itemize}
\itemindent=-18pt\itemsep=0pt\parsep=0pt\vskip 1cm

\item[] Ademollo M \etal\ 1976a \pl\ B {\bf 62} 105
\item[] Ademollo M \etal\ 1976b \np\ B {\bf 111} 77
\item[] Aldazabal G, Hussain F and Zhang R 1987 \pl\ B {\bf 185} 89
\item[] Alvarez-Gaum\'{e} L and Freedman D Z 1981 \cmp\ {\bf 80} 443
\item[] Alvarez-Gaum\'{e} L, Freedman D Z and Mukhi S 1981
\ap\ {\bf 134} 85
\item[] Antoniadis I, Bachas C, Kounnas C and Windey P 1986
\pl\ B {\bf 171} 51
\item[] Atick J J and Witten E 1988 \np\ B {\bf 310} 291
\item[] Bandos I A, Sorokin D, Tonin M, Pasti P and Volkov D V
1995 \np B {\bf 446} 79
\item[] Banks T and Dixon L J 1988 \np\ B {\bf 307} 93
\item[] Banks T, Dixon L J, Friedan D and Martinec E 1988
\np\ B {\bf 299} 613
\item[] Banks T, Nemeschansky D and Sen A 1986 \np\ B {\bf 277} 67
\item[] Becker K, Becker M and Strominger S 1995 \np\ B {\bf 456} 130
\item[] Blencowe M P and Duff M J 1988 \np\ B {\bf 310} 387
\item[] Bluhm R, Dolan L and Goddard P 1987 \np\ B {\bf 289} 364
\item[] Bonneau G and Valent G 1994 \cqg\ {\bf 11} 1133
\item[] Braden H W 1987 \np\ B {\bf 291} 516
\item[] Callan C G, Friedan D, Martinec E J and Perry M J 1985
\np\ B {\bf 262} 593
\item[] Coleman S 1975 \pr\ D {\bf 11} 2088
\item[] D'Adda A and Lizzi F 1987 \pl\ B {\bf 191} 85
\item[] Delduc F, Kalitzin F and Sokatchev E 1990
\cqg\ {\bf 7} 1567
\item[] Dine M and Seiberg N 1986 \pl\ B {\bf 180} 364
\item[] Dirac P A M (1967) {\it Lectures on Quantum Mechanics}\
(New York: Academic Press Inc.)
\item[] Dixon L J, Kaplunovsky V S and Vafa C 1987 \np\ B {\bf 294} 43
\item[] Eguchi T, Ooguri H, Taormina A and Yang S-K 1989
\np\ B {\bf 193}.
\item[] Fradkin E S and Tseytlin A A 1981 \pl\ B {\bf 106} 63
\item[] \dash 1985a \pl\ B {\bf 162} 295
\item[] \dash 1985b \np\ B {\bf 261} 1
\item[] Gates S J, Howe P S and Hull C M 1989 \pl\ B {\bf 227} 49
\item[] Goddard P, Nahm W and Olive D 1985 \pl\ B {\bf 160} 111
\item[] Goddard P, Kent A and Olive D 1986 \cmp\ {\bf 103} 105
\item[] Goddard P and Olive D 1986 \ijmp\ A {\bf 1} 303
\item[] Green M B 1987 \np\ B {\bf 293} 593
\item[] \dash 1994 \pl\ B {\bf 329} 435
\item[] Greene B P, Morrison D R and Strominger A 1995
\np\ B {\bf 451} 109
\item[] Gross D J, Harvey J A, Martinec E and Rohm R 1985
\np\ B {\bf 256} 253
\item[] \dash 1986
\np\ B {\bf 267} 75
\item[] Gross D J and Mende P F 1987 \pl\ B {\bf 197} 129
\item[] \dash 1988 \np\ B {\bf 303} 407
\item[] Halpern M 1975 \pr\ D {\bf 12} 1684
\item[] Hanson A, Regge T and Teitelboim C 1976 {\it Constrained
Hamiltonian Systems}\ (Rome: Accademia Nazionale dei Lincei)
\item[] Harvey J A and Strominger A 1995 \np\ B {\bf 449} 535;
erratum \np\ B {\bf 458} 456
\item[] Henneaux M and Teitelboim C 1992 {\it Quantization of Gauge
Systems}\ (Princeton, NJ: Princeton University Press)
\item[] Howe P S and Papadopoulos G 1988 \cqg\ {\bf 5} 1647
\item[] Hull C M 1986a \pl\ B {\bf 178} 357
\item[] \dash 1986b \np\ B {\bf 267} 266
\item[] \dash 1994 \mpl\ A {\bf 9} 161
\item[] \dash 1995 String dynamics at strong coupling
{\it Preprint} QMW-95-50, hep-th/9512181
\item[] Hull C M, Papadopoulos G and Spence B 1991 \np\ {\bf 363} 593
\item[] Hull C M and Spence B 1989 \pl\ B {\bf 232} 204
\item[] \dash 1990 \np\ B {\bf 345} 493
\item[] \dash 1991 \np\ B {\bf 353} 379
\item[] Hull C M and Townsend P K 1986 \pl\ B {\bf 178} 187
\item[] Hull C M and Witten E 1985 \pl\ B {\bf 160} 398
\item[] Jack I, Jones D R T, Mohammedi N and Osborn H 1990
\np\ B {\bf 332} 359
\item[] Kawai H, Lewellyn D C and Tye S-H H 1986a \pr\ D {\bf 34} 3794
\item[] \dash 1986b \prl\ {\bf 57} 1832
\item[] \dash 1987a \np\ B {\bf 288} 1
\item[] \dash 1987b \pl\ B {\bf 191} 63
\item[] Kutasov D and Martinec E 1996 New Principles
for String/Membrane Unification, {\it Preprint} 
EFI-96-04, hep-th/9602049
\item[] Kutasov D, Martinec E and O'Loughlin M 1996
Vacua of M-theory and N=2 Strings, {\it Preprint} EFI-96-07,
hep-th/9603116
\item[] Losev A, Moore G, Nekrasov N and Shatashvili S 1995
Four-Dimensional Avatars of Two-Dimensional RCFT, {\it Talk presented
at the USC Strings '95 conference},
{\it Preprint} PUPT-1564,ITEP-TH.5/95,YCTP-P15/95,
hep-th/9509151
\item[] Lovelace C 1986 \np\ B {\bf 273} 413
\item[] Mathur S and Mukhi S 1987 \pr\ D {\bf 36} 465
\item[] \dash 1988 \np\ B {\bf 302} 130
\item[] Moore G and Nelson P 1984 \prl\ {\bf 53} 1519
\item[] \dash 1985 \cmp\ {\bf 100} 83
\item[] Morrison D R and Vafa C 1996 Compactifications of F-Theory
on Calabi-Yau Threefolds I {\it Preprint} DUKE-TH-96-106, HUTP-96-A007
hep-th/9602114; Compactifications of F-Theory on Calabi-Yau Threefolds II,
{\it Preprint} DUKE-TH-96-107, HUTP-96-A012, hep-th/9603161
\item[] Nair V P and Schiff J 1990 \pl\ B {\bf 246} 423
\item[] \dash 1992 \np\ B {\bf 371} 329
\item[] Ooguri H and Vafa C 1990 \mpl\ A {\bf 5} 1389
\item[] \dash 1991a \np\ B {\bf 361} 469
\item[] \dash 1991b \np\ B {\bf 367} 83
\item[] Pierce D M 1986 A (1,2) Heterotic String With
Gauge Symmetry, {\it Preprint} IFP-604-UNC, hep-th/9601125
\item[] Polchinski J 1988 \pl\ B {\bf 209}, 252
\item[] \dash 1994 What is String Theory? {\it Lectures
from the 1994 Les Houches summer school}, Fluctuating geometries
in statistical mechanics and field theory, {\it Preprint}
NSF-ITP-94-97, hep-th/9411028
\item[] R\v{o}cek M and Verlinde E 1992 \np\ B {\bf 373} 630
\item[] Schwarz J A 1995a \pl\ {\bf 360} 13; erratum \pl\ B {\bf 364} 252
\item[] \dash 1995b Superstring dualities {\it Preprint}
CALT-68-2019, hep-th/9509148
\item[] \dash 1995c \pl\ B {\bf 367} 97
\item[] Schwimmer A and Seiberg N 1987 \pl\ B {\bf 184} 191
\item[] Sen A 1985 \prl\ {\bf 55} 1846, \pr\ D {\bf 32} 2102
\item[] \dash 1986 \pl\ B {\bf 166} 300; 
\pl\ B {\bf 174} 277; \np\ B {\bf 278} 289
\item[] \dash 1995 \np\ B {\bf 450} 103
\item[] Shenker S H 1995 Another length scale in string theory?
{\it Preprint} RU-95-53, hep-th/9509132
\item[] Townsend P K 1995 P-Brane Democracy {\it Preprint}
DAMPT-95-34, hep-th/9507048
\item[] Tseytlin A A 1996 Self-duality of Born-Infeld action
and Dirichlet 3-brane of type $IIB$ superstring theory
{\it Preprint} Imperial/TP/95-96/26, hep-th/9602064
\item[] Vafa C 1996 Evidence for F-Theory {\it Preprint}
HUTP-96-A004, hep-th/9602022
\item[] van Nieuwenhuizen P 1987 {\it Santiago 1987,
Proceedings, Quantum mechanics of fundamental systems}
(New York: Plenum) 
\item[] Windey P 1986 \cmp\ {\bf 105} 511
\item[] Witten E 1988 \prl\ {\bf 61} 670
\item[] \dash 1996a Phase Transitions in M Theory and F Theory
{\it Preprint} IASSNS-HEP-96-26, hep-th/9603150
\item[] \dash 1996b Nonperturbative Superpotentials in String Theory
{\it Preprint} IASSNS-HEP-96-29, hep-th/9604030

\end{itemize}

\end{document}